\documentstyle[preprint,eqsecnum,aps]{revtex}

\begin{document}
\draft
\preprint{HEP/123-qed}
\title{Polarization transfer observables for quasielastic proton-nucleus
scattering in terms of a complete Lorentz invariant representation of the NN
scattering matrix}

\author{B.I.S. van der Ventel, G.C. Hillhouse and P.R. De Kock}
\address{University of Stellenbosch, Stellenbosch, 7600, South Africa
}
\author{S.J. Wallace}
\address{Department of Physics, University of Maryland, College Park, \\
Maryland, 20742
}
\date{\today}
\maketitle
\begin{abstract}
For the calculation of 
polarization transfer observables for quasielastic scattering of
protons on nuclei, a formalism in the context of the Relativistic Plane Wave
Impulse Approximation is developed, in which the interaction matrix is expanded
in terms of a complete set of 44 independent invariant amplitudes. 
A boson-exchange model is used to predict the 39 amplitudes which were 
omitted in the 
formerly used five-term parameterization
(the SPVAT form) of the nucleon-nucleon scattering matrix. Use of the complete
set of amplitudes eliminates the arbitrariness of the five-term representation.
\end{abstract}

\narrowtext

\section{Introduction}
\label{section_introduction}

Quasielastic scattering of protons on nuclei is an attractive phenomenon for
the study of the basic nucleon-nucleon interaction in the nuclear medium,
because it exhibits the 
approximate behavior of the scattering of a nucleon on only one nucleon of
the target nucleus. Quasielastic scattering has been modeled by the 
Relativistic Plane Wave Impulse Approximation \cite{HorowitzMurdock88}
which considers it as a single-step process, whereby the projectile interacts
with only one nucleon of the target nucleus, while the rest of the nucleons
remain inert. 
The well known and outstanding success of the original RPWIA
was its prediction of the analyzing power for 
$ ^{40} {\rm Ca} ( \vec{p}, \vec{p}^{ \, \, '} ) $ and 
$ ^{208} {\rm Pb} ( \vec{p}, \vec{p}^{ \, \, '} ) $ at 500 MeV; a case in which
all non-relativistic models failed \cite{hillhouse_PhD98}.

In the RPWIA approach, the description of the initial and final free
particle in the medium is based on mean-field theory, as described 
by Serot and Walecka
in Ref.
\cite{serot86}. In the RPWIA model 
the associated Dirac plane waves have their
free nucleon mass decreased by the real part of the 
average nuclear scalar field to yield an effective
nucleon mass. The values of the effective masses serve
as an indicator of the nuclear medium effects on the NN interaction. 

In former theoretical studies of scattering 
\cite{McNielShepardWallacePRLa,McNielShepardWallacePRLb,HorowitzMurdock88}
the nucleon-nucleon scattering matrix $ ( \hat{F} ) $
was parameterized in terms of the five Fermi covariants
which is commonly referred to as the SPVAT form of $ \hat{F} $ or the IA1
model. It should be stressed, however, that eventhough the SPVAT form gave
reasonable results for elastic and quasileastic scattering observables,
it is in principle not correct, 
since as was first pointed out in
Ref. \cite{AdamsBleszynski84}, a five-term representation of the relativistic
NN scattering matrix is necessarily ambiguous. In addition, Tjon and Wallace
\cite{TjonWallace85} have shown that a general Lorentz invariant representation
of $ \hat{F} $ (referred to as the IA2 model) contains additional 
terms that cannot be neglected. The IA2 representation of $ \hat{F} $
contains, in fact, 44 independent invariant amplitudes, instead of the 
previously used five, which are 
consistent with parity and time-reversal invariance as
well as charge symmetry together with the on-mass-shell condition for the 
external nucleons.
Comparison to the limited data available with subsequent and more refined 
calculations 
\cite{hillhouse_PhD98,hillhouse94,hillhouse95,hillhouse98}
have also revealed that quasielastic $ (\vec{p}, \vec{p}^{\, \, '} ) $ and
$ ( \vec{p}, \vec{n} ) $ prefer different five-term representations of 
$ \hat{F} $: The $ (\vec{p}, \vec{n} ) $ data favor a pseudovector 
$ \pi NN $ coupling, whereas the $ (\vec{p}, \vec{p}^{\, \, '} ) $ data
are consistent with a pseudoscalar term for the $ \pi NN $ vertex.
Therefore, the most basic question which has to be addressed is 
the representation of the NN scattering matrix.

In the current application of the RPWIA to quasielastic scattering, 
the following components play a key role:
\begin{enumerate}
 \item The amplitudes in the basic two-nucleon interaction, 
which are partly determined from free NN
scattering data and partly from a solution of the Bethe-Salpeter equation 
employing a meson-exchange model for the NN force.
 \item The Lorentz covariant set constructed from the Dirac matrices which
serves as a representation for $ \hat{F} $.
 \item The effective nucleon mass for both projectile and target nucleons 
interacting in the nuclear
medium.
\end{enumerate}

In this paper a theoretical formalism is presented for
the calculation of calculate polarization transfer 
observables for quasielastic proton-nucleus scattering using a general
Lorentz invariant representation of the NN scattering matrix 
\cite{TjonWallace87}; a systematic survey of the predictive power of the model
compared to data will be presented in a future paper. 
By adhering to the symplifying features of the RPWIA, one
can focus on the basic NN interaction without introducing additional 
complications.
A complete and unambiguous expansion of $ \hat{F} $ allows for a 
correct incorporation of medium effects (within the RPWIA framework) 
and therefore one can distinguish between
experimental results which are genuine medium effects and those which
arise due to other effects not taken into account by the RPWIA.
In Section \ref{section_RPWIA} we briefly review the RPWIA and
also discuss the ambiguities of the SPVAT form of $ \hat{F} $.
In Section \ref{section_IA2_rep_NNscatmat} the general
Lorentz invariant representation of $ \hat{F} $ is discussed. 
Section \ref{section_transf_inv_amps_to_effec_amps} presents the
transformation from invariant amplitudes to effective amplitudes, while
in Section \ref{section_spin_obs_effec_amps} expressions for the 
spin observables
are derived in terms of the effective amplitudes.
A calculation of complete sets of spin observables, based on the IA2 model,
for quasielastic $^{40}$Ca$( \vec{p}, \vec{p}^{\; '} )$ scattering at
500 MeV is presented in Section \ref{section_results}. 
Section \ref{section_summary} summarizes the main aspects of this paper.

\section{Relativistic Plane Wave Impulse Approximation}
\label{section_RPWIA}

Complete sets of spin observables 
\footnote{The spin observables are defined in
Section \ref{section_spin_obs_effec_amps}}
($ P $, $ A_{y} $, $ D_{l' l} $, $ D_{s' s} $, $ D_{s' l} $,
$ D_{l' s} $, $ D_{n n} $) for quasielastic 
$ (\vec{p}, \vec{p}^{\, \, '}) $ and $ (\vec{p}, \vec{n} ) $ scattering
are calculated within a relativistic framework using the Relativistic
Plane Wave Impulse Approximation (RPWIA) \cite{HorowitzMurdock88}. 
The RPWIA models
quasielastic scattering as a single-step process whereby the projectile knocks
out a single bound nucleon from the nucleus. The rest of the nucleons are 
assumed 
to remain inert, but their effect is taken into account in that the 
free mass of the
projectile and target nucleons are shifted to {\em effective masses}, 
$ M_{1} $ and $ M_{2} $ respectively. 
In the context of the Walecka model \cite{serot86}
the effective masses can both be
calculated microscopically as follows: For the projectile,
\begin{eqnarray*}
 M_{1} & = & M + <S(\vec{r} \, \, )>
\end{eqnarray*}
where $ M $ is the free nucleon mass,
$ <S(\vec{r} \, \, )> $ is the average value over the whole nucleus 
of the real part of the scalar
potential found by weighting it with $ T (b ) $, the transmission probability
through the nucleus at an impact parameter $ b $, 
and with the nuclear density
$ \rho ( r ) $ for the specific nucleus 
\cite{hillhouse94}. The effective
mass of the target nucleon is determined from
\begin{eqnarray*}
 M_{2} & = & M - g_{s} < \phi ( \vec{r} \, \, ) >
\end{eqnarray*}
where $ g_{s} $ is the scalar meson coupling constant and 
$ \phi ( \vec{r} \, \, ) $
is the scalar field for the specific nucleus 
with the averaging done as described
above. Values of $ M_{1} $ and $ M_{2} $ for specific nuclei and incident 
laboratory energies can be found in Table II of Ref. \cite{hillhouse94}.
Experimental data seem to suggest that
the spin observables are target independent \cite{chan89,wakasaPhD96}, 
and therefore we
assume, as a first step, a Fermi-gas approximation for the target nucleus. 
The RPWIA therefore reduces quasielastic proton-nucleus 
scattering to a two-body 
scattering process with Dirac spinors (containing effective nucleon masses)
describing the external nucleons. A graphical representation of the scattering
process is depicted in Fig. \ref{two-body_scattering_process}.

Refering to Fig. \ref{two-body_scattering_process}, 
the projectile Dirac spinor is given by:

\begin{eqnarray}
 \label{projectilespinor}
   U ( \vec{p}_{1}, M_{1}, s_{i} ) 
   & = &
   \left[
         \frac{E_{1}^{*} + M_{1}}{2 M_{1}}
   \right]^{\frac{1}{2}} \,
   \left(
         \begin{array}{c}
           \phi ( s_{i} ) 
           \\
           \frac{ \vec{\sigma} \cdot \vec{p}_{1}}{ E_{1}^{*} + M_{1}}
           \phi ( s_{i} )
         \end{array}
   \right)
\end{eqnarray}
where $ E_{1}^{*^{2}} \, = \, \vec{p}_{1}^{\, \, 2} \, + 
\, M_{1}^{2} $ and the
spinor is normalized to 
$
  \bar{U} ( \vec{p}_{1}, M_{1}, s ) \, 
   U ( \vec{p}_{1}, M_{1}, s' ) \, = \, \delta_{s s'}.
$ 
Similar expressions exist 
for the other three spinors labelled by $ \vec{p}_{2} $,
$ \vec{k}_{1} $ and $ \vec{k}_{2} $.
The following four-momenta are also
defined:
\begin{eqnarray*}
 p_{1}^{*} \, = \, ( E_{1}^{*}, \vec{p}_{1} ), 
 & \hspace{1cm} &
 p_{2}^{*} \, = \, ( E_{2}^{*}, \vec{p}_{2} ), 
 \\ 
 k_{1}^{*} \, = \, ( E_{1}^{*'}, \vec{k}_{1} ), 
 & &
 k_{2}^{*} \, = \, ( E_{2}^{*'}, \vec{k}_{2} ) 
\end{eqnarray*}
where $ p_{i}^{*^{2}} \, = \, M_{i}^{2} $ and 
$ k_{i}^{*^{2}} \, = \, M_{i}^{2} $ ($ i \, = \, 1, \, 2 $). 
For handling the polarization, one requires the spin projection operator,
\begin{eqnarray}
 \label{spin_proj_op_1}
  P ( \hat{n} ) & = & \frac{1}{2} ( I_{2} + \vec{\sigma} \cdot \hat{n} )
\end{eqnarray}
for the direction $ \hat{n} $ where $ I_{2} $ is the $ 2 \times 2 $ unit 
matrix. In the basis of Pauli spinors, $ \phi(\hat{n}) $ for spin direction
$ \hat{n} $, we have
\begin{eqnarray}
 \label{spin_proj_op_2}
  P ( \hat{n} ) & = & \phi ( \hat{n} ) \phi^{\dagger} ( \hat{n} ).
\end{eqnarray} 
Defining
\begin{eqnarray*}
  \bar{U} ( \vec{p}_{1}, M_{1}, s_{i} ) 
   & = &
  U^{ \dagger} ( \vec{p}_{1}, M_{1}, s_{i} ) \,
  \gamma^{0} 
\end{eqnarray*}
where the convention of Ref. \cite{BjorkenDrellRelQM} 
is used for the gamma-matrices,
the Lorentz invariant matrix element for the scattering process depicted in 
Fig. \ref{two-body_scattering_process} is given by:
\begin{eqnarray}
  \label{invme}
  {\cal M} & = &
  [\bar{U} ( \vec{k}_{1}, M_{1}, s_{f} ) \otimes
        \bar{U} ( \vec{k}_{2}, M_{2}, s^{'}_{2} ) 
  ] \, \hat{F} \,
  [
        U ( \vec{p}_{1}, M_{1}, s_{i} ) \otimes
        U ( \vec{p}_{2}, M_{2}, s_{2} ) 
  ]
\end{eqnarray}
where
$ \hat{F} $ is the $ 16 \times 16 $ nucleon-nucleon scattering matrix.
The question arises as to what form of $ \hat{F} $ is to be used in 
Eq. (\ref{invme}) assuming parity and time-reversal 
invariance as well as charge symmetry.
Once a choice of $ \hat{F} $ has been made, analytical expressions for all spin
observables, namely the unpolarized double differential cross section, 
the analyzing power and the
polarization transfer observables can obtained from Eq. (\ref{invme}).
All previous calculations 
\cite{HorowitzMurdock88,HorowitzIqbal86,hillhouse94,hillhouse95,hillhouse98}
of spin observables for quasielastic proton-nucleus scattering have been
parameterized $ \hat{F} $ in terms of only the five Fermi covariants:
\begin{eqnarray}
 \nonumber
  \hat{F} & = &
  F_{S} ( I_{4} \otimes I_{4} ) \, + \, F_{P} ( \gamma^{5} \otimes \gamma^{5} )
  \, + \, F_{V} ( \gamma^{\mu} \otimes \gamma_{\mu} ) \, + \,
  \\
  & &  
  \label{NNscatmatSPVAT}
  F_{A} ( \gamma^{5} \gamma^{\mu} \otimes \gamma^{5} \gamma_{\mu} ) \, + \,
  F_{T} ( \sigma^{\mu \nu} \otimes \sigma_{\mu \nu} )
\end{eqnarray}  
where the latter is commonly called 
the SPVAT form or IA1 representation of
$ \hat{F} $. The amplitudes, $ F_{L} $ $( L = S,P,V,A,T )$ are
obtained by fitting to free NN scattering data \cite{HorowitzIqbal86}. 
{\em This procedure, however, does not uniquely fix the form of the 
matrix $ \hat{F} $ }. To see this we note that the pseudoscalar covariant,
$ PS \, = \, \gamma^{5} \otimes \gamma^{5} $, has exactly the same matrix
elements between positive energy free mass Dirac spinors 
$ ( M_{1} = M_{2} = M ) $
as the pseudovector covariant,
$ PV \, = \, \frac{q \hspace*{-1.20mm} / \hspace*{1.75mm} 
\gamma^{5}}{2 M} \otimes
\frac{q \hspace*{-1.20mm} / \hspace*{1.75mm} \gamma^{5}}{2 M} $, i.e.,
\begin{eqnarray*}
  \left[
        \bar{U}_{1} ( M ) \otimes \bar{U}_{2} ( M ) 
  \right]
  \left[
        PV - PS 
  \right]
  \left[
        U_{1} ( M ) \otimes U_{2} ( M )
  \right]
  & = & 0.
\end{eqnarray*}
This is called the equivalence theorem \cite{hamilton59}.
We can therefore replace $ PS $ with $ PV $ in 
Eq. (\ref{NNscatmatSPVAT}) without
altering the amplitudes, $ F_{L} $. Eventhough these two representations are
equivalent on-shell $( p^{2} \; = \; M^{2} )$
they will give different results when sandwiched between
{\em positive energy Dirac spinors containing an effective nucleon mass}, 
since then matrix
elements between negative energy states now also enter. This is because the 
effective mass spinor can always be expanded in a free mass basis:
\begin{eqnarray*}
   U ( \vec{p}_{1}, M_{1}, s_{i} ) 
   & = &
   \alpha_{U} U ( \vec{p}_{1}, M, s_{i} ) \, + \,
   \alpha_{V} V ( \vec{p}_{1}, M, s_{i} )
\end{eqnarray*}
where $ V $ is the negative energy Dirac spinor \cite{BjorkenDrellRelQM}.
There also exists the relation \cite{hillhouse94},
\begin{eqnarray}
 \label{PV_PS_me}
  {\cal M}_{PV} 
  & = &
  \frac{M_{1} M_{2}}{M^{2}} {\cal M}_{PS}
\end{eqnarray}  
where $ {\cal M}_{PS} $ and $ {\cal M}_{PV} $ are the contribution of the
pseudoscalar covariant and pseudovector covariant respectively to the
invariant matrix element given by Eq. (\ref{invme}). 
Note that in Eq. (\ref{PV_PS_me}), the pseudovector covariant is
$ PV \, = \, \frac{q \hspace*{-1.20mm} / \hspace*{1.75mm} 
\gamma^{5}}{2 M} \otimes
\frac{q \hspace*{-1.20mm} / \hspace*{1.75mm} \gamma^{5}}{2 M} $, but where
$ q \, = \, p_{1}^{*} - k_{1}^{*} \, = \, k_{2}^{*} - p_{2}^{*} $, i.e., the
momenta are on-mass-shell with respect to the effective masses, $ M_{1} $ and
$ M_{2} $. In the equivalence theorem, the momenta must be on-mass-shell with
respect to the free mass.
The above equality has been used in 
Refs. \cite{hillhouse94,hillhouse95,hillhouse98} to investigate the sensitivity of
the spin observables to the difference between using a pseudoscalar covariant
or a pseudovector covariant.
The ambiguity which is inherent in any five-term or incomplete representation
of $ \hat{F} $ (such as the IA1 representation) was first pointed out in
Ref. \cite{AdamsBleszynski84}.
Tjon and Wallace have developed a
general Lorentz invariant representation of $ \hat{F} $. The formalism
can be found in Refs. \cite{TjonWallace85,TjonWallace87} and is applied to 
{\em elastic} proton-nucleus scattering in Refs.
\cite{TjonWallacePRL85,TjonWallacePRC85,TjonWallacePRC87}.
We will refer to this as the IA2 representation of 
$ \hat{F} $
and discuss, in the next section, its application to 
{\em quasielastic} proton-nucleus scattering. 

\section{IA2 representation of $ \hat{F} $ applied to quasielastic
proton-nucleus scattering}
\label{section_IA2_rep_NNscatmat}

From Eqs (3.1) and (3.18) in Ref. \cite{TjonWallacePRC87} 
the IA2 representation of
$ \hat{F} $ is given by:
\begin{eqnarray}
 \label{NNscatmatIA2}
   \hat{F} & = &
   \sum_{\rho_{1} \rho'_{1}; \rho_{2} \rho'_{2}} \,
   \sum_{n = 1}^{13} \, F_{n}^{\rho_{1} \rho'_{1} \rho_{2} \rho'_{2}} \,
   [
     \Lambda_{\rho'_{1}} ( \vec{k}_{1}; M ) \otimes
     \Lambda_{\rho'_{2}} ( \vec{k}_{2}; M )
   ] \, K_{n} \,
   \nonumber
   \\
   & &
   [
    \Lambda_{\rho_{1}} ( \vec{p}_{1}; M ) \otimes
    \Lambda_{\rho_{2}} ( \vec{p}_{2}; M )
   ]
\end{eqnarray}
where $ M $ refers to the free nucleon mass.
Henceforth the notation
$$
  \{ \rho \} \, = \, \rho_{1} \rho'_{1}; \rho_{2} \rho'_{2}
$$
will be used.
In Eq. (\ref{NNscatmatIA2}), 
$ F_{n}^{\{ \rho \}} $ ($ n \, = \, 1 - 13) $
are the invariant amplitudes for each rho-spin sector (which is defined by
the rho-spin labels,  $ \rho_{1} \rho'_{1}; \rho_{2} \rho'_{2} $, where 
$ \rho \, = \, \pm $ ), $ \Lambda_{\rho} ( \vec{p}, M ) $ 
are covariant projection
operators given by:
\begin{eqnarray}
 \label{rho_spin_projection_operator}
  \Lambda_{\rho} ( \vec{p}, M ) 
  & \; \; = \; \; &
  \frac{\rho p \hspace*{-1.80mm} / \hspace*{1.75mm} + M }{2 M}
  \nonumber
  \\
  & \; \; = \; \; &
  \frac{\rho ( E \gamma^{0} \, - \, \vec{p} \cdot \vec{\gamma}) + M }{2 M}
\end{eqnarray}
where $ E^{2} \, = \, \vec{p}^{\, \, 2} \, + \, M^{2} $, and 
$ K_{n} $ $ ( n \, = \, 1-13 ) $ are kinematic covariants constructed 
from the Dirac matrices:
\begin{eqnarray*}
  K_{1} & = & S \, = \, I_{4} \otimes I_{4}
  \\
  K_{2} & = & P \, = \, \gamma^{5} \otimes \gamma^{5}
  \\
  K_{3} & = & V \, = \, \gamma^{\mu} \otimes \gamma_{\mu}
  \\
  K_{4} & = & A \, = \, \gamma^{5} \gamma^{\mu} \otimes \gamma^{5} \gamma_{\mu}
  \\
  K_{5} & = & T \, = \, \sigma^{ \mu \nu} \otimes \sigma_{\mu \nu}
  \\
  K_{6} & = & Q_{11, \mu} ( I_{4} \otimes \gamma^{\mu} )
  \\
  K_{7} & = & Q_{22, \mu} ( \gamma^{\mu} \otimes I_{4} )
  \\
  K_{8} & = & Q_{11, \mu} ( \gamma^{5} \otimes \gamma^{5} \gamma^{\mu} )
  \\
  K_{9} & = & Q_{22, \mu} ( \gamma^{5} \otimes \gamma^{5} \gamma^{\mu} ) 
  \\
  K_{10} & = & Q_{12, \mu} ( I_{4} \otimes \gamma^{\mu} ) \, \tilde{S}
  \\
  K_{11} & = & Q_{21, \mu} ( \gamma^{\mu} \otimes I_{4} ) \, \tilde{S}
  \\
  K_{12} & = & Q_{12, \mu} ( \gamma^{5} \otimes \gamma^{5} \gamma^{\mu} ) \,
  \tilde{S}
  \\
  K_{13} & = & Q_{21, \mu} ( \gamma^{5} \gamma^{\mu} \otimes \gamma^{5} ) \,
  \tilde{S}
\end{eqnarray*}
where 
\begin{eqnarray*}
  Q_{i j, \mu} & = &
  \frac{ ( p'_{i} \, + \, p_{j} )_{\mu}}{2 M} \qquad {\rm with} \qquad
  p'_{1} \, = \, k_{1} \, \, {\rm and } \, \,
  p'_{2} \, = \, k_{2}.
\end{eqnarray*}
With each combination of rho-spin labels \{  $ \rho_{1} \rho'_{1} \rho_{2}
\rho'_{2}  $ \} is associated a pair $ ( ij ) $ to index a specific rho-spin sector 
(or subclass): See Table I of \cite{TjonWallacePRC87}. For example
$
  \{ ++++ \} \, \equiv \, ( 11 )
$
and
$
  \{ +-+- \} \, \equiv \, ( 22 ).
$
Parity and time-reversal invariance, together with charge symmetry and the
on-mass-shell condition for external nucleons, lead to $ \hat{F} $ being
completely  specified by 44 independent invariant amplitudes 
\cite{TjonWallace87}. Five amplitudes in subclass $ \hat{F}^{11} $ 
are completely
specified by fitting to 
physical free NN scattering data and are therefore identical
to the SPVAT amplitudes in the IA1 
representation of $ \hat{F} $. 
The remaining 39 off-shell amplitudes (contained in subclasses
$ \hat{F}^{12} $ to $ \hat{F}^{44} $) are obtained by solving the 
Bethe-Salpeter equation in a three-dimensional quasi-potential reduction
\cite{vanFaassenTjon83,vanFaassenTjon84}, with pure pseudovector pion-nucleon
coupling,
to determine a complete set of helicity
amplitudes. The invariant amplitudes are related via matrix equations to the 
helicity amplitudes \cite{TjonWallace87}.
The IA2 representation is a 
complete and unambiguous expansion of $ \hat{F} $ since covariants cannot
be added or changed arbitrarily without 
violating the above-mentioned symmetries. Amplitudes
which are solely determined by physical scattering data are isolated in subclass
$ \hat{F}^{11} $ while the remaining amplitudes are determined by solving
a dynamical equation, the Bethe-Salpeter equation using a meson-exchange model
for the NN force.

From Eq. (\ref{NNscatmatIA2}) four cases concerning the combination of 
projectile and target nucleon masses can be 
distinguished:
\begin{enumerate}
 \item No medium effect $( M_{1} = M_{2} = M )$: In this case only subclass
$ \hat{F}^{11} $ will contribute to the invariant scattering amplitude.
 \item Projectile Relativity $( M_{1} \neq M; M_{2} = M )$: Contributions
to the invariant scattering amplitude
arise from $ \hat{F}^{11}, \hat{F}^{21}, \hat{F}^{31} $ and 
$\hat{F}^{41} $ where the latter three subclasses require {\em at least}
projectile relativity for a contribution.
 \item Target Relativity $(M_{1} = M; M_{2} \neq M )$: Contributions 
to the invariant scattering amplitude arise
from $ \hat{F}^{11}, \hat{F}^{12}, \hat{F}^{13} $ and $ \hat{F}^{14} $
where the latter three subclasses require {\em at least} target 
relativity for a contribution.
 \item Target and projectile relativity $( M_{1} \neq M; M_{2} \neq M )$:
 Now all subclasses will contribute to the invariant scattering 
amplitude but 
    $ \hat{F}^{22} $,  $ \hat{F}^{23} $, 
    $ \hat{F}^{24} $, $ \hat{F}^{32} $,
    $ \hat{F}^{33} $, $ \hat{F}^{34} $, $ \hat{F}^{42} $, $ \hat{F}^{43} $ and
    $ \hat{F}^{44} $ require {\em at least } projectile {\em and} target
    relativity for a contribution.
\end{enumerate}
From Eq. (\ref{NNscatmatIA2}) we see 
that medium effects can never occur in subclass $ \hat{F}^{11} $ due
to the accompanying positive energy projection operators. 
Medium effects in the IA2 representation of $ \hat{F} $ arise
only due to off-shell amplitudes (which are contained in the subclasses
$ \hat{F}^{12} $ to $ \hat{F}^{44} $). This is in contrast to IA1 where
medium effects are included only in subclass $ \hat{F}^{11} $.
One can now substitute Eq. (\ref{NNscatmatIA2}) into Eq. (\ref{invme}) and
proceed from there to calculate the spin observables in terms of 
$ | {\cal M} |^{2} $ 
which is directly related to the invariant
amplitudes $ F_{n}^{\{\rho\}} $. We will, however, not follow this 
direct approach
due to the following reasons:
\begin{enumerate}
 \item Following the standard procedure 
(see Ref. \cite{BjorkenDrellRelQM} for example) one
 finds that $ | {\cal M} |^{2} $ contains traces over at least eight
 gamma matrices. The number of gamma matrices increase as the covariants
 become more complicated. Since the number of terms generated by such a trace 
 is given by 
 $ \displaystyle{\frac{N !}
  { ( \frac{N}{2}) ! \, 2^{\frac{N}{2}}}} $ (where $ N $
 refers to the number of gamma matrices), and since there 
 is a double sum over the 
 rho-spin sectors, a very large number of terms will occur.
 \item Since we are applying a relativistic formalism to a Nuclear Physics
 problem, it might be more instructive to rewrite the NN scattering matrix in
 a form which is more familiar to traditional Nuclear Physics.
\end{enumerate}
We will therefore follow a similar approach as in 
Ref. \cite{FurnstahlWallace93}
where an {\em effective $ t $-matrix is derived} 
which is a $ 4 \times 4 $ 
matrix, but which still contains all the information coming from the 
relativistic analysis.
From Eq. (\ref{projectilespinor}) we can write:
\begin{eqnarray*}
 U ( \vec{p}_{1}, M_{1}, s_{i} ) 
 & = &
 \left( 
       \frac{E_{1}^{*}}{M_{1}} 
 \right)^{\frac{1}{2}} 
 u^{+} ( \vec{p}_{1}, M_{1} ) 
  \phi ( s_{i} )
\end{eqnarray*}
where, as a $ 4 \times 2 $ matrix:
\begin{eqnarray}
 \label{pos_distorted_spinor}
  u^{+} ( \vec{p}_{1}, M_{1} ) 
  & = &
 \left(
       \frac{E_{1}^{*} + M_{1}}{2 E_{1}^{*}}
 \right)^{\frac{1}{2}} \,
 \left(
       \begin{array}{c}
         I_{2}
         \\
\displaystyle{\frac{ \vec{\sigma} \cdot \vec{p}_{1}}{E_{1}^{*} + M_{1}}}
       \end{array}
 \right).
\end{eqnarray}
Similarly 
\begin{eqnarray*}
 \bar{U} ( \vec{p}_{1}, M_{1}, s_{i} ) 
 & = &
 \left(
       \frac{E_{1}^{*}}{M_{1}}
 \right)^{\frac{1}{2}} \,
 \phi^{ \dagger} ( s_{i} ) \bar{u}^{+} ( \vec{p}_{1}, M_{1} )
\end{eqnarray*}
where, as a $ 2 \times 4 $ matrix:
\begin{eqnarray*}
  \bar{u}^{+} ( \vec{p}_{1}, M_{1} ) 
  & = &
  u^{+^{\dagger}} ( \vec{p}_{1}, M_{1} ) \gamma^{0}.
\end{eqnarray*}
$ u^{ \rho} ( \vec{p} \, \, ) $ (where $ \rho \, = \, \pm $) contains no
reference to the spin and is normalized to
\begin{eqnarray*}
 u^{\rho '^{\dagger}} ( \rho ' \vec{p} \, ) u^{\rho} ( \rho \vec{p} \, )
 & = &
 \delta_{\rho ' \rho}.
\end{eqnarray*}
In terms of $ u^{+} $ 
the invariant matrix element, (Eq. \ref{invme}) is given by 
\begin{eqnarray*}
 {\cal M} & = &
 \left(
  \frac{E_{1}^{*} E_{2}^{*} E_{1}^{*^{'}} 
        E_{2}^{*^{'}}}{M_{1}^{2} M_{2}^{2}}
 \right)^{\frac{1}{2}}
 [
  \phi^{\dagger} ( s_{f} ) \bar{u}^{+} ( \vec{k}_{1}, M_{1} ) \otimes
  \phi^{\dagger} ( s_{2}^{'} ) \bar{u}^{+} ( \vec{k}_{2}, M_{2} ) 
 ] \, \hat{F} \,
 \\
 & &
 [
  \phi ( s_{i} ) u^{+} ( \vec{p}_{1}, M_{1} ) \otimes
  \phi ( s_{2} ) u^{+} ( \vec{p}_{2}, M_{2} )
 ].
\end{eqnarray*}
Use of the identity
$
  {\rm A C} \otimes {\rm B D}
  =
  ( {\rm A} \otimes {\rm B} ) ( {\rm C} \otimes {\rm D} ) 
$, where $ \otimes $ refers to the usual Kronecker product,
leads to the expression
\begin{eqnarray}
 \label{invmepauliIA2}
  {\cal M} & = &
  \left(
    \frac{E_{1}^{*} E_{2}^{*} E_{1}^{*^{'}} 
         E_{2}^{*^{'}}}{M_{1}^{2} M_{2}^{2}}
  \right)^{\frac{1}{2}}
  [
   \phi^{\dagger} ( s_{f} ) \otimes \phi^{\dagger} ( s_{2}^{'} )
  ] \,
  [
   \bar{u}^{+} ( \vec{k}_{1}, M_{1} ) \otimes 
   \bar{u}^{+} ( \vec{k}_{2}, M_{2} ) 
  ] \hat{F}
  \nonumber
  \\
  & &
  [
   {u}^{+} ( \vec{p}_{1}, M_{1} ) \otimes 
   {u}^{+} ( \vec{p}_{2}, M_{2} ) 
  ] 
  [
   \phi ( s_{i} ) \otimes \phi ( s_{2} )
  ]. 
\end{eqnarray}
Defining the {\em effective $ t $-matrix as}:
\begin{eqnarray}
 \label{effective_tmatrix}
  \hat{t} & = &
  [
   \bar{u}^{+} ( \vec{k}_{1}, M_{1} ) \otimes 
   \bar{u}^{+} ( \vec{k}_{2}, M_{2} ) 
  ] \hat{F}
  [
   {u}^{+} ( \vec{p}_{1}, M_{1} ) \otimes 
   {u}^{+} ( \vec{p}_{2}, M_{2} ) 
  ]
\end{eqnarray}
and
$ g_{1} = [ (E_{1}^{*} E_{2}^{*} E_{1}^{*^{'}} E_{2}^{*^{'}} ]/
M_{1}^{2} M_{2}^{2} ]^{\frac{1}{2}} $,
Eq. (\ref{invmepauliIA2}) becomes:
\begin{eqnarray}
 \label{invmetmatrix}
  {\cal M} & = &
  g_{1} \,
  [
   \phi^{\dagger} ( s_{f} ) \otimes \phi^{\dagger} ( s_{2}^{'} )
  ] \hat{t}
  [
   \phi ( s_{i} ) \otimes \phi ( s_{2} )
  ]. 
\end{eqnarray}
Since $ \hat{t} $ is a $ 4 \times 4 $ matrix it can be expanded in terms of
a basis constructed from the Pauli matrices and 
the momenta of the scattering 
process. Define 
the three-momentum transfer 
$
\vec{q} \, = \, \vec{p}_{1} \, - \, \vec{k}_{1} \, = \,
\vec{k}_{2} \, - \, \vec{p}_{2}, 
$ 
the average momentum
$
\vec{p}_{a} \, = \, \frac{1}{2} ( \vec{p}_{1} \, + \, \vec{k}_{1} )
$ 
and a vector orthogonal to both $ \vec{q} $ and $ \vec{p}_{a} $, 
$
\vec{N} \, = \, \vec{q} \, \times \, \vec{p}_{a} \, 
= \, \vec{p}_{1} \, \times \, \vec{k}_{1}.
$
Note that 
\begin{eqnarray*}
 \vec{q} \cdot \vec{p}_{a} 
 & = &
 \frac{1}{2} ( \vec{p}_{1}^{\, \, 2} \, - \, \vec{k}_{1}^{ \, \, 2} ).
\end{eqnarray*}
For quasielastic scattering $ | \vec{p}_{1} | $ $ \neq $ 
$ | \vec{k}_{1} | $ and therefore $ \vec{q} $ and $ \vec{p}_{a} $ are not
orthogonal, however, $ \vec{N} \cdot \vec{q} \, = \, \vec{N} \cdot \vec{p}_{a} 
\, = 0 $.
{\em Assuming only parity invariance, $ \hat{t} $ can be written in terms of a
set of eight linearly independent matrices in the spin of the two interacting
nucleons:}
\begin{eqnarray}
 \label{texpand_pauli_2}
  \hat{t} & = &
  \sum_{n=1}^{8} \, b_{n} \, ( \chi^{(1)}_{n} \otimes \chi^{(2)}_{n} ) 
\end{eqnarray}
where
$$
\begin{array}{cc}
 \chi_{1}^{(1)} \, = \, I_{2}, &
 \chi_{1}^{(2)} \, = \, I_{2}.
 \\
 \chi_{2}^{(1)} \, = \, \vec{N} \cdot
 \vec{\sigma}, & \chi_{2}^{(2)} \, = \, \vec{N} \cdot \vec{\sigma}.
 \\
 \chi_{3}^{(1)} \, = \, \vec{N} \cdot
 \vec{\sigma}, & \chi_{3}^{(2)} \, = \, I_{2}.
 \\
 \chi_{4}^{(1)} \, = \, I_{2}, &
 \chi_{4}^{(2)} \, = \, \vec{N} \cdot \vec{\sigma}.
 \\
 \chi_{5}^{(1)} \, = \, \vec{q} \cdot 
 \vec{\sigma}, & \chi_{5}^{(2)} \, = \, \vec{q} \cdot \vec{\sigma}.
 \\
 \chi_{6}^{(1)} \, = \, 
 \vec{p_{a}} \cdot 
 \vec{\sigma}, & \chi_{6}^{(2)} \, = \, \vec{p_{a}} \cdot \vec{\sigma}.
 \\
 \chi_{7}^{(1)} \, = \, \vec{q} \cdot 
 \vec{\sigma}, & \chi_{7}^{(2)} \, = \, \vec{p_{a}} \cdot \vec{\sigma}.
 \\
 \chi_{8}^{(1)} \, = \, \vec{p_{a}} \cdot 
 \vec{\sigma}, & \chi_{8}^{(2)} \, = \, \vec{q} \cdot \vec{\sigma}.
\end{array}
$$
In the next section the invariant amplitudes, $ F_{n}^{\rho} $
$(n = 1 - 13)$ are transformed to a set of eight effective amplitudes
$ b_{n} $ $(n = 1 - 8)$, and expressions for the spin observables are
derived in terms of the effective amplitudes.

\section{ 
Transformation from the invariant amplitudes to the effective amplitudes.}
\label{section_transf_inv_amps_to_effec_amps}

Expressions for the effective amplitudes of $ \hat{t} $ are now derived: 
Taking the trace of Eq. (\ref{texpand_pauli_2}) yields
\begin{eqnarray}
 \label{eq_a1}
  b_{1} & = & \frac{1}{4} {\rm Tr} [ \, \hat{t} \, ].
\end{eqnarray}
Multiply Eq. (\ref{texpand_pauli_2}) with 
$ ( \vec{N} \cdot \vec{\sigma} \otimes \vec{N} \cdot \vec{\sigma} ) $
and take the trace of the resulting equation.
Since $ \vec{N} \cdot \vec{q} \, = \, \vec{N} \cdot \vec{p_{a}} \, = \, 0 $
there will be no contribution from the last four terms of 
Eq. (\ref{texpand_pauli_2}) and therefore
\begin{eqnarray}
 \label{eq_a2}
  b_{2} & = & 
  \frac{1}{4 \vec{N}^{\, \, 4}} Y_{2} ( \vec{N}, \vec{N} )
\end{eqnarray}
where
\begin{eqnarray*}
 Y_{2} ( \vec{a}, \vec{b} ) 
 & = &
 {\rm Tr} [( \vec{a} \cdot \vec{\sigma} \otimes 
 \vec{b} \cdot \vec{\sigma} ) \hat{t} \, ].
\end{eqnarray*}
Similar arguments lead to 
\begin{eqnarray}
 \label{eq_a3}
  b_{3} & = & \frac{1}{4 \vec{N}^{\, \, 2}}
  {\rm Tr} [ (\vec{N} \cdot \vec{\sigma} \otimes I_{2} ) \hat{t} \, ]
\end{eqnarray}
and 
\begin{eqnarray}
 \label{eq_a4}
  b_{4} & = & \frac{1}{4 \vec{N}^{\, \, 2}} 
  {\rm Tr} [(I_{2} \otimes \vec{N} \cdot \vec{\sigma} ) \hat{t} \, ].
\end{eqnarray}
Following the same reasoning as above, one can also derive a set of
four coupled equations relating the amplitudes $ b_{5} $, $ b_{6} $, 
$ b_{7} $ and $ b_{8} $. A set of coupled equations arise since the 
vectors $ \vec{q} $ and $ \vec{p}_{a} $ {\em are not orthogonal for
quasielastic scattering}, 
(i.e. $ \vec{q} \cdot \vec{p}_{a} \, \neq \, 0 $ ).
The solutions are: 
\begin{eqnarray}
 \label{eq_a5}
  b_{5} & = &
  \frac{r_{2}^{\, \, 2} Y_{2} ( \vec{p}_{a}, \vec{p}_{a} ) \, - \,
  r_{2} r_{3} Y_{2} ( \vec{p}_{a}, \vec{q} \; ) \, - \,
  r_{2} r_{3} Y_{2} ( \vec{q}, \vec{p}_{a} ) \, + \,
  r_{3}^{\, \, 2} Y_{2} ( \vec{q}, \vec{q} \; )}
  {M^{2} ( r_{1} r_{3} \, - \, r_{2}^{\, \, 2} )^{2}}
  \\
 \label{eq_a6}
  b_{6} & = &
  \frac{r_{1}^{\, \, 2} Y_{2} ( \vec{p}_{a}, \vec{p}_{a} ) \, - \,
  r_{1} r_{2} Y_{2} ( \vec{p}_{a}, \vec{q} \; ) \, - \,
  r_{1} r_{2} Y_{2} ( \vec{q}, \vec{p}_{a} ) \, + \,
  r_{2}^{\, \, 2} Y_{2} ( \vec{q}, \vec{q} \; )}
  {M^{2} ( r_{1} r_{3} \, - \, r_{2}^{\, \, 2} )^{2}}
  \\
 \label{eq_a7}
  b_{7} & = &
  \frac{-r_{1} r_{2} Y_{2} ( \vec{p}_{a}, \vec{p}_{a} ) \, + \,
  r_{2}^{\, \, 2} Y_{2} ( \vec{p}_{a}, \vec{q} \; ) \, + \,
  r_{1} r_{3} Y_{2} ( \vec{q}, \vec{p}_{a} ) \, - \,
  r_{2} r_{3} Y_{2} ( \vec{q}, \vec{q} \; ) }
  {M^{2} ( r_{1} r_{3} \, - \, r_{2}^{\, \, 2} )^{2}}
  \\
 \label{eq_a8}
  b_{8} & = &
  \frac{-r_{1} r_{2} Y_{2} ( \vec{p}_{a}, \vec{p}_{a} ) \, + \,
  r_{1} r_{3} Y_{2} ( \vec{p}_{a}, \vec{q} \; ) \, + \,
  r_{2}^{\, \, 2} Y_{2} ( \vec{q}, \vec{p}_{a} ) \, - \,
  r_{2} r_{3} Y_{2} ( \vec{q}, \vec{q} \; ) }
  {M^{2} ( r_{1} r_{3} \, - \, r_{2}^{\, \, 2} )^{2}}
\end{eqnarray}
where
\begin{eqnarray*}
 r_{1} & = & \frac{2 \vec{q}^{\, \, 2}}{M},
 \\
 r_{2} & = & \frac{2 \vec{p}_{a} \cdot \vec{q}}{M} \qquad {\rm and}
 \\
 r_{3} & = & \frac{2 \vec{p}_{a}^{ \, \, 2}}{M}.
\end{eqnarray*}
The next step is to derive an expression for the $ \hat{t} $-matrix 
which is convenient for use in the calculation of the traces
which determine the effective amplitudes. 
Substitution of  Eq. (\ref{NNscatmatIA2}) into Eq. (\ref{effective_tmatrix}) 
leads to
\begin{eqnarray}
 \label{texpand_pauli_3}
  \hat{t} & = &
  \sum_{\{ \rho \}} \sum_{n = 1}^{13} \, F_{n}^{\{ \rho \}} \, 
  \left[ 
   \bar{\Gamma}_{\rho_{1}^{'}} ( \vec{k}_{1}, M, M_{1} ) \otimes
   \bar{\Gamma}_{\rho_{2}^{'}} ( \vec{k}_{2}, M, M_{2} ) 
  \right] K_{n}
 \nonumber
 \\
 & &
\left[ 
   \Gamma_{\rho_{1}} ( \vec{p}_{1}, M, M_{1} ) \otimes
   \Gamma_{\rho_{2}} ( \vec{p}_{2}, M, M_{2} ) 
\right] 
\end{eqnarray}
where we have introduced the 
$ 4 \times 2 $ $ \Gamma $-matrices defined as:
\begin{eqnarray}
 \label{gamma_func_1}
  \Gamma_{ \rho } ( \vec{p}, M, M^{*} ) 
  & = &
  \Lambda_{ \rho } ( \vec{p}, M ) \,
  u^{+} ( \vec{p}, M^{*} ) .
\end{eqnarray}
In Eq. (\ref{gamma_func_1}) $ M^{*} $ denotes an effective mass and 
Eq. (\ref{pos_distorted_spinor}) has been generalized to:
\begin{eqnarray}
 \label{pos_distorted_spinor_general}
  u^{+} ( \vec{p}, M^{*} ) 
  & = &
  \left(
        \begin{array}{c}
         I_{2} \, \phi ( \vec{p}_{1}, M^{*} )
         \\
         \vec{\sigma} \cdot \vec{p} \, \chi ( \vec{p}, M^{*} )
        \end{array}
  \right).
\end{eqnarray}
Eq. (\ref{pos_distorted_spinor_general}) reduces to
Eq. (\ref{pos_distorted_spinor}) if we set
\begin{eqnarray*}
  \phi ( \vec{p}_{1}, M_{1} ) 
  & = &
  \left(
        \frac{ E^{*} ( \vec{p}_{1} ) \, + \, M_{1} }
        {2 E^{*} ( \vec{p}_{1} ) }
  \right)^{\frac{1}{2}}
\end{eqnarray*}
where $ E^{*} ( \vec{p}_{1} ) \, = \, \sqrt{\vec{p}_{1}^{\, \, 2} \, + \,
M_{1}^{2} } $  and 
\begin{eqnarray*}
  \chi ( \vec{p}_{1}, M_{1} )
  & = &
  \phi ( \vec{p}_{1}, M_{1} ) \,
  \left(
        E^{*} ( \vec{p}_{1} ) \, + \, M_{1} ) 
  \right)^{-1}.
\end{eqnarray*}
We can obtain an explicit expression for $ \Gamma_{\rho} $ as follows:
From Eq. (\ref{pos_distorted_spinor_general}) we can write
\begin{eqnarray*}
 u^{+} ( \vec{p}, M^{*} )
 & = &
 \left(
       \begin{array}{c}
        I_{2} \, \phi ( \vec{p}, M^{*} ) 
        \\
        0
       \end{array}
 \right) \, + \,
 \left(
       \begin{array}{c}
        0
        \\
        \vec{\sigma} \cdot \vec{p} \, \chi ( \vec{p}, M^{*} )
       \end{array}
 \right)
\end{eqnarray*}
and therefore
\begin{eqnarray}
 \label{pos_distorted_spinor_e-basis}
  u^{+} ( \vec{p}, M^{*} )
  & = &
  \phi ( \vec{p}, M^{*} ) \, ( \hat{e}_{1} \otimes I_{2} ) \, + \,
  \chi ( \vec{p}, M^{*} ) \, 
  ( \hat{e}_{2} \otimes \vec{\sigma} \cdot \vec{p} \, \, )
\end{eqnarray}
where
\begin{eqnarray*}
 \hat{e}_{1} \, = \, 
 \left(
       \begin{array}{c}
        1 \\ 0
       \end{array}
 \right)
 & {\rm and } & \hspace{1cm}
 \hat{e}_{2} \, = \,
 \left(
       \begin{array}{c}
        0 \\ 1
       \end{array}
 \right)
\end{eqnarray*}
with $ \hat{e}_{i}^{\dagger} \hat{e}_{j} \, = \, \delta_{i j} $.
To write the $ \rho $-spin projection operator in 
$ (2 \times 2) \otimes (2 \times 2) $ form we recall that
\begin{eqnarray}
 \label{gamma_0_pauli_form}
  \gamma^{0} & = &
  \left(
        \begin{array}{cc}
         I_{2} & 0
         \\
         0 & -I_{2}
        \end{array}
  \right) \, = \,
  \sigma_{3} \otimes I_{2} \qquad {\rm and}
  \\
 \label{gamma_i_pauli_form}
  \vec{p} \cdot \vec{\gamma} 
  & = & 
  \left(
        \begin{array}{cc}
         0 & \vec{p} \cdot \vec{\sigma}
         \\
         -\vec{p} \cdot \vec{\sigma} & 0
        \end{array}
  \right) \, = \,
  i \sigma_{2} \otimes \vec{p} \cdot \vec{\sigma}.
\end{eqnarray}
Substitution of Eqs. (\ref{gamma_0_pauli_form}) and 
(\ref{gamma_i_pauli_form}) into 
Eq. (\ref{rho_spin_projection_operator})
leads to 
\begin{eqnarray}
 \label{gamma_func_e-basis}
  \Lambda_{\rho} ( \vec{p}, M )
  & = &
  \frac{\rho E_{p}}{2 M} ( \sigma_{3} \otimes I_{2} ) \, - \,
  \frac{i \rho}{2 M} ( \sigma_{2} \otimes \vec{p} \cdot \vec{\sigma} )
  \, + \, \frac{1}{2} ( I_{2} \otimes I_{2} ).
\end{eqnarray}
Substitution of Eqs. (\ref{pos_distorted_spinor_e-basis}) and
(\ref{gamma_func_e-basis}) into Eq. (\ref{gamma_func_1}), and 
using the properties of the
Pauli matrices, allows one to write:
\begin{eqnarray}
\label{gamma_func_2}
  \Gamma_{\rho} ( \vec{p}, M, M^{*} )
  & = &
  \sum_{i=1}^{2} \, h_{\rho}^{(i)} ( \vec{p}, M, M^{*} ) \,
  \left[ \hat{e}_{i} \otimes A_{i} ( \vec{p} \, \, ) \right]
\end{eqnarray}
where
\begin{eqnarray*}
  A_{i} ( \vec{p} \, \, ) & = &
  \left \{
          \begin{array}{c}
           I_{2}; \qquad i \, = \, 1 
           \\
           \vec{p} \cdot \vec{\sigma}; \qquad i \, = \, 2
          \end{array}
  \right. 
\end{eqnarray*}
with
\begin{eqnarray*}
  h_{\rho}^{(1)} ( \vec{p}, M, M^{*} ) 
  & = &
  \frac{\rho E_{p}}{2 M} \phi ( \vec{p}, M^{*} ) \, - \,
  \frac{\rho \vec{p}^{\, \, 2}}{2 M} \chi ( \vec{p}, M^{*} ) \, + \,
  \frac{1}{2} \phi ( \vec{p}, M^{*} ) \qquad {\rm and}
\\
  h_{\rho}^{(2)} ( \vec{p}, M, M^{*} )
  & = &
  \frac{\rho}{2 M} \phi ( \vec{p}, M^{*} ) \, - \,
  \frac{ \rho E_{p}}{2 M} \chi ( \vec{p}, M^{*} ) \, + \,
  \frac{1}{2} \chi ( \vec{p}, M^{*} ).
\end{eqnarray*}
Similar steps lead to 
\begin{eqnarray}
\label{gamma_bar_func_e-basis}
  \bar{\Gamma}_{\rho} ( \vec{p}, M, M^{*} )
  & = &
  \sum_{i=1}^{2} \, j_{\rho}^{(i)} ( \vec{p}, M, M^{*} ) \, 
  \left[ \hat{e}_{i}^{\dagger} \otimes A_{i} ( \vec{p} \, ) \right]
\end{eqnarray}
where
\begin{eqnarray*}
  j_{\rho}^{(1)} ( \vec{p}, M, M^{*} )
  & = & 
  h_{\rho}^{(1)} ( \vec{p}, M, M^{*} ) \qquad {\rm and}
  \\
  j_{\rho}^{(2)} ( \vec{p}, M, M^{*} )
  & = & 
  -h_{\rho}^{(2)} ( \vec{p}, M, M^{*} ).
\end{eqnarray*}
To calculate the effective amplitudes, the contribution of each covariant 
to the trace relations must be determined. 
$ \hat{t} $ is calculated from Eq. (\ref{texpand_pauli_3}) using the
explicit forms of $ \Gamma_{\rho_{i}} ( \vec{p}, M, M^{*} ) $ and
$ \bar{\Gamma}_{\rho_{i}} ( \vec{p}, M, M^{*} ) $ in
Eqs. (\ref{gamma_func_2}) and (\ref{gamma_bar_func_e-basis}).
Use is then made of Eqs. (\ref{eq_a1}) to (\ref{eq_a8}) to determine
the effective amplitudes, $ b_{1} $ to $ b_{8} $.
This procedure requires the calculation of traces of a set of matrices
$ t_{i} $, where $ i \, = \, 1 - 46 $. With each covariant
is associated a set of t-matrices of which the trace must be taken.
We wrote a program in the MATHEMATICA computer language
to do the required trace algebra. 
The effective amplitudes are linear functions of 
$ F_{n}^{\{ \rho \}} $, i.e.
\begin{eqnarray}
 \label{amps_effec_inv}
  b_{i} & = & b_{i} ( \{ F_{n}^{\{ \rho \} } \}) \qquad {\rm where} \, \, \, 
  i \, = \, 1 - 8.
\end{eqnarray}
The isospin zero (isospin one) effective amplitudes are obtained by
substituting the isospin zero (isospin one) invariant amplitudes
into Eq. (\ref{amps_effec_inv}).

\section{Expressions for spin observables in terms of the 
effective amplitudes}
\label{section_spin_obs_effec_amps}

In this section expressions are derived for the unpolarized cross section,
the analyzing power and the polarization transfer observables in terms
of the effective amplitudes $ b_{n} $ for both
$ ( \vec{p}, \vec{p}^{\, \, '} ) $ and
$ ( \vec{p}, \vec{n} ) $ scattering.

Working in the nucleon-nucleon
laboratory frame, the spin in the incident beam direction is
described in terms of three orthogonal unit vectors
$ ( \hat{l}, \hat{s}, \hat{n} ) $, where $ \hat{l} $ is along the beam 
direction, $ \hat{s} $ lies perpendicular and to the side of $ \hat{l} $
in the scattering plane, and the normal unit vector is
$ \hat{n} \, = \, \hat{l} \times \hat{s} $. Similarly the spin of the final
beam is described in terms of $ ( \hat{l}^{\, '}, \hat{s}^{\, '},
\hat{n} ) $. 

\begin{enumerate}
 \item \underline{Unpolarized double differential cross section}

For the scattering process in Fig. \ref{two-body_scattering_process}
one can write down the
expression for the differential cross section, $ d \sigma $
\cite{BjorkenDrellRelQM}:
\begin{eqnarray*}
  d \sigma & = & \frac{1}{ | \vec{v}_{1} - \vec{v}_{2} | }
  \left(
        \frac{M_{1}}{E_{1}^{*}} \frac{M_{2}}{E_{2}^{*}}
  \right)
  \left(
        \frac{M_{1}}{E_{1}^{*^{'}}} \frac{M_{2}}{E_{2}^{*^{'}}}
  \right)
  2 \pi^{4} \delta ( p_{1}^{*} + p_{2}^{*} - k_{1}^{*} - k_{2}^{*} )
  \\
  \nonumber
  & &
  \frac{d^{3} \vec{k}_{1}}{ ( 2 \pi )^{3}}
  \frac{d^{3} \vec{k}_{2}}{ ( 2 \pi )^{3}}
  | {\cal M} |^{2}.
\end{eqnarray*}
As we consider the quasielastic scattering at energies much higher than the
interaction energies amongst the target nucleons, we assume the latter to be
practically non-interacting. Therefore, the momentum distribution of these
nucleons can be obtained in a Fermi-gas model.
Following the same arguments as in Ref. \cite{HorowitzMurdock88}
allows one to write down
the following expression for the 
{\em double differential cross section:}
\begin{eqnarray*}
  \frac{d \sigma }{d \Omega_{1}^{'} d E_{1}^{'} }
  & = &
  \frac{ | \vec{k}_{1} | E_{1}^{'} }{ | \vec{q} | E_{1}^{*^{'}}}
  \int_{p_{min}}^{k_{f}} \, d | \vec{p}_{2} | \, \, d \phi \, \, 
  | \vec{p}_{2} | f ( p_{1}^{*}, p_{2}^{*}, k_{f} )  | {\cal M} |^{2}
\end{eqnarray*}
where
\begin{eqnarray}
\label{f-function_in_xsection}
  f ( p_{1}^{*}, p_{2}^{*}, k_{f} )
  & = &
  \frac{3}{16 \pi^{3} k_{f}^{3}}
  \frac{M_{1}^{2} M_{2}^{2}}
  { [ 
     ( p_{1}^{*} \cdot p_{2}^{*} - M_{1}^{2} M_{2}^{2} )^{2}
    ] ^{\frac{1}{2}}} 
\end{eqnarray}
and
\begin{eqnarray}
 \label{eq_p_min}
 p_{min} & = &
 \left| 
       \frac{q}{2} - \frac{\omega^{*}}{2} 
       \left[
             1 - \frac{4 M^{2}_{2}}{ q_{\mu} q^{\mu}}
       \right]^{\frac{1}{2}} 
 \right|.
\end{eqnarray}
In Eq. (\ref{eq_p_min}), $ q^{\mu} $ is the four-momentum transfer 
$ q^{\mu} \; = \; ( \omega^{*}, \vec{q} \; ) $ where
\begin{eqnarray*}
 \omega^{*} \, = \, E^{*}_{1} - E_{1}^{*^{'}} \, = \, 
 E_{2}^{*^{'}} - E_{2}
 & \qquad {\rm and} \qquad &
 \vec{q} \, = \, \vec{p}_{1} - \vec{k}_{1} \, = \,
 \vec{k}_{2} - \vec{p}_{2}.
\end{eqnarray*}
To obtain the Fermi momentum $ k_{f} $, the required effective density is
calculated in an eikonal approximation as shown in 
Ref. \cite{HorowitzMurdock88}.
More refined values of $ k_{f} $ for specific target nuclei can be 
found in Table II of Ref. \cite{hillhouse94}.
Define the function
\begin{eqnarray*}
  \Gamma^{''} ( \vec{p}_{1}, \vec{p}_{2}, \vec{k}_{1}, \vec{k}_{2} )
  & = &
  \sum_{s_{i}, s_{f}} \, \sum_{s_{2}, s_{2}^{'}} \,
  | {\cal M} |^{2}.
\end{eqnarray*}
Substitution of Eq. (\ref{texpand_pauli_2}) into Eq. (\ref{invmetmatrix})
leads to 
\begin{eqnarray*}
 {\cal M} & = &
 g_{1} \sum_{n = 1}^{8} \, b_{n} 
 \left[
       \phi^{\dagger} ( s_{f} ) \chi_{n}^{(1)} \phi ( s_{i} )
 \right]
 \left[
       \phi^{\dagger} ( s_{2}^{'} ) \chi_{n}^{(2)} \phi ( s_{2} )
 \right]
\end{eqnarray*}
and therefore one can write
\begin{eqnarray*}
  \Gamma^{''} ( \vec{p}_{1}, \vec{p}_{2}, \vec{k}_{1},
  \vec{k}_{2} ) 
  & = &
  g_{1}^{2} \sum_{m,n = 1}^{8} \, b_{m}^{*} b_{n} \, 
  {\rm Tr} [ \chi_{n}^{(1)} \chi_{m}^{(1)} ]
  {\rm Tr} [ \chi_{n}^{(2)} \chi_{m}^{(2)} ].
\end{eqnarray*}
An explicit expression for 
$ \Gamma^{''} ( \vec{p}_{1}, \vec{p}_{2}, \vec{k}_{1}, \vec{k}_{2} ) $
is given in the appendix.
To obtain the unpolarized cross section, one sums over the initial spin and 
average over the final spin which leads to:
\begin{eqnarray}
  \nonumber
  \left(
        \frac{d \sigma}{d \Omega_{1}^{'} d E_{1}^{'}} 
  \right)_{{\rm unpol}}
  & = &
  \frac{ | \vec{k}_{1} | E_{1}^{'} }{ 4 | \vec{q} \, | E_{1}^{*^{'}}}
  \int_{p_{min}}^{k_{f}} \, d | \vec{p}_{2} | \, \, d \phi \, \,
  | \vec{p}_{2} | f ( p_{1}^{*}, p_{2}^{*}, k_{f} ) 
  \\
 \label{unpol_cross_section}
  & &
  \Gamma^{''} ( \vec{p}_{1}, \cdots
  \vec{k}_{2} ). 
\end{eqnarray}
For $ ( \vec{p}, \vec{p}^{\, \, '} ) $ scattering
\begin{eqnarray}
  \nonumber
  \left(
        \frac{d \sigma}{d \Omega_{1}^{'} d E_{1}^{'}} 
  \right)_{{\rm unpol}}^{(p,p')}
  & = &
  \frac{ | \vec{k}_{1} | E_{1}^{'} }{ 4 | \vec{q} \, | E_{1}^{*^{'}}}
  \int_{p_{min}}^{k_{f}} \, d | \vec{p}_{2} | \, \, d \phi \, \,
  | \vec{p}_{2} | f ( p_{1}^{*}, p_{2}^{*}, k_{f} ) 
  \\
  \nonumber
  & &
  \, \, ( Z_{eff} \Gamma^{''} ( \vec{p}_{1}, \cdots
  \vec{k}_{2}, \{ b_{i} ( I = 1 ) \} ) +
  N_{eff} \Gamma^{''} ( \vec{p}_{1}, \cdots
  \vec{k}_{2}, \{ b_{i}^{ave} \} ) 
\end{eqnarray}
where 
$$
  b_{i}^{ave} \, = \, \frac{1}{2} 
  (b_{i} ( I = 0 ) \, + \, b_{i} ( I = 1 ) ).
$$
For the charge-exchange reaction $ ( \vec{p}, \vec{n} ) $,
\begin{eqnarray*}
 \nonumber
  \left(
        \frac{d \sigma}{d \Omega_{1}^{'} d E_{1}^{'}} 
  \right)_{{\rm unpol}}^{(p,n)}
  & = &
  \frac{ | \vec{k}_{1} | E_{1}^{'} }{ 4 | \vec{q} \, | E_{1}^{*^{'}}}
  \int_{p_{min}}^{k_{f}} \, d | \vec{p}_{2} | \, \, d \phi \, \,
  | \vec{p}_{2} | f ( p_{1}^{*}, p_{2}^{*}, k_{f} )  
  \\
  & &
  N_{eff} \Gamma^{''} ( \vec{p}_{1}, \cdots
  \vec{k}_{2}, \{ b_{i}^{ch-ex} \} ) 
\end{eqnarray*}
where the charge-exchange amplitudes are defined as:
\begin{eqnarray*}
 b_{i}^{ch-ex} & = & \frac{1}{2} [ b_{i} ( I = 1 ) - b_{i} ( I = 0 ) ].
\end{eqnarray*}
The quantities $ Z_{eff} $ and $ N_{eff} $ are defined in 
Ref. \cite{HorowitzMurdock88} and values for specific targets are given
in Table II of Ref. \cite{hillhouse94}.
 \item \underline{Analyzing power}

The definition of the analyzing power is given in terms of polarized
double differential cross sections as:
\begin{eqnarray}
 \label{analyzing_power_def}
  A_{y} & = &
  \frac{ \frac{d \sigma}{ d \Omega_{1}^{'} 
  d E_{1}^{'}} ( \hat{s}_{f} = +\hat{n} ) -
  \frac{ d \sigma}{ d \Omega_{1}^{'} d E_{1}^{'}}
  ( \hat{s}_{f} = -\hat{n} ) }
  { \frac{d \sigma}{ d \Omega_{1}^{'} 
  d E_{1}^{'}} ( \hat{s}_{f} = +\hat{n} ) + 
  \frac{ d \sigma}{ d \Omega_{1}^{'} d E_{1}^{'}}
  ( \hat{s}_{f} = -\hat{n} ) }
\end{eqnarray}
where, for example,
\begin{eqnarray*}
  \frac{d \sigma }{d \Omega_{1}^{'} d E_{1}^{'} }
  ( \hat{s}_{f} )
  & = &
  \frac{ | \vec{k}_{1} | E_{1}^{'} }{ | \vec{q} \, | E_{1}^{*^{'}}}
  \int_{p_{min}}^{k_{f}} \, d | \vec{p}_{2} | \, \, d \phi \, \, 
  | \vec{p}_{2} | f ( p_{1}^{*}, p_{2}^{*}, k_{f} )  
  \frac{1}{2} \tilde{\Gamma}^{'} ( \hat{s}_{f} )
\end{eqnarray*}
is averaged over incident spin directions $ \hat{s}_{i} $, and the target
particles' initial and final spin as contained in the factor:
\begin{eqnarray}
  \nonumber
  \tilde{\Gamma}^{'} ( \vec{p}_{1},..., \vec{k}_{2}, \hat{s}_{f} )
  & = &
  \sum_{s_{i}, s_{2}, s_{2}^{'}} \, 
  | {\cal M} |^{2}
  \\
 \label{big_gamma_tilde_prime_2}
  & = &
  {\rm Tr} [ \chi^{(1)}_{m} \hat{P} ( \hat{s}_{f} ) \chi^{(1)}_{n} ]
  {\rm Tr} [ \chi^{(2)}_{m} \chi^{(2)}_{n} ]
\end{eqnarray}
where use was made of Eqs. (\ref{spin_proj_op_1}) and 
(\ref{spin_proj_op_2}).
A calculation of the traces in Eq. (\ref{big_gamma_tilde_prime_2})
shows that 
$ \tilde{\Gamma}^{'} ( \vec{p}_{1}, ..., \hat{s}_{f} ) $ has the 
following structure:
\begin{eqnarray}
  \nonumber
  \tilde{\Gamma}^{'} ( \vec{p}_{1}, ..., \hat{s}_{f} )
  & = &
  f_{1} ( \vec{p}_{1}, ..., \vec{k}_{2} ) \, + \, 
  f_{2} ( \vec{p}_{1}, ..., \vec{k}_{2} ) \vec{N} \cdot \hat{s}_{f} 
  \, + \, 
  \\
 \label{big_gamma_tilde_prime_3}
  & &
  f_{3} ( \vec{p}_{1}, ..., \vec{k}_{2} ) 
  \vec{p}_{a} \cdot ( \vec{q} \times \hat{s}_{f} ).
\end{eqnarray}
Defining the combination function
\begin{eqnarray}
\label{big_gamma_prime_1}
  \Gamma^{'} ( \vec{p}_{1}, ..., \vec{k}_{2}, \hat{s}_{f} )
  & = &
  \tilde{\Gamma}^{'} ( \vec{p}_{1}, ..., \vec{k}_{2}, \hat{s}_{f} )
  \, - \, 
  \tilde{\Gamma}^{'} ( \vec{p}_{1}, ..., \vec{k}_{2}, -\hat{s}_{f} ).
\end{eqnarray}
and using Eq. (\ref{big_gamma_tilde_prime_3}) 
in Eq. (\ref{big_gamma_prime_1})
yields
\begin{eqnarray*}
  \Gamma^{'} ( \hat{s}_{f} ) 
  & = &
  2 ( f_{2} ( \vec{p}_{1}, ..., \vec{k}_{2} ) \vec{N} \cdot \hat{s}_{f}
  \, + \,
  f_{3} ( \vec{p}_{1}, ..., \vec{k}_{2} ) 
  \vec{p}_{a} \cdot ( \vec{q} \times \hat{s}_{f} ).
\end{eqnarray*}
The explicit forms of the functions $ f_{2} $
and $ f_{3} $ can be inferred 
from Eq. (\ref{big_gamma_prime_xxx}) 
in the appendix. 
If $ \hat{s}_{f} \, = \, \hat{n} $ then
$$
  \vec{N} \cdot \hat{n} \, = \, p_{1} k_{1} \sin \theta_{L}
$$
and
$$
  \vec{p}_{a} \cdot ( \vec{q} \times \hat{n} ) \, = \,
  -p_{1} k_{1} \sin \theta_{L}.
$$
The analyzing power for the $ ( \vec{p}, \vec{p}^{ \, \, '} ) $ reaction
is given by
{\small
\begin{eqnarray*}
  A_{y} ( \vec{p}, \vec{p}^{\, \, '} )
  & = &
  \frac{ \int_{p_{min}}^{k_{f}} 
  d | \vec{p}_{2} | \, \, d \phi | \, \, \vec{p}_{2} |
  f ( p_{1}^{*}, .., p_{2}^{*}, k_{f} ) \,
  ( Z_{eff} \Gamma^{'} ( \hat{n}, \{ b_{i} ( I = 1 ) \} ) \, + \, 
  N_{eff} \Gamma^{'} ( \hat{n}, \{ b_{i}^{ave} \} ) )}
  { \int_{p_{min}}^{k_{f}} 
  d | \vec{p}_{2} | \, \, d \phi | \, \, \vec{p}_{2} |
  f ( k_{f} ) \,
  ( Z_{eff} \Gamma^{''} 
  ( \{ b_{i} ( I = 1 ) \} ) \, + \, 
  N_{eff} \Gamma^{''} 
  ( \{ b_{i}^{ave} \} ) )}
\end{eqnarray*}
}
and the analyzing power for the $ ( \vec{p}, \vec{n} ) $ reaction is
given by
\begin{eqnarray}
 \label{analyzing_power_pn}
  A_{y} ( \vec{p}, \vec{n} ) 
  & = &
  \frac{ \int_{p_{min}}^{k_{f}} d | \vec{p}_{2} | d \phi | \vec{p}_{2} |
  f ( k_{f} ) \,
  \Gamma^{'} ( \hat{n}, \{ b_{i}^{ch-ex} \} ) }
  { \int_{p_{min}}^{k_{f}} d | \vec{p}_{2} | d \phi | \vec{p}_{2} |
  f ( k_{f} ) \,
  \Gamma^{''} 
  ( \{ b_{i}^{ch-ex} \} ) }.
\end{eqnarray}
Since a $ (\vec{p}, \vec{n} \, ) $ reaction implies that the incident proton
could only have scattered off a neutron, 
we set $ Z_{eff} \, = \, 0 $ and therefore
$ N_{eff} $ appears as a common factor in the numerator and denominator and
cancels out, which means that $ N_{eff} $ 
does not appear in 
Eq. (\ref{analyzing_power_pn}).
 \item \underline{Polarization transfer observables}

The polarization transfer observables are defined in terms of linear 
combinations
of polarized double differential cross sections as follows:
\begin{eqnarray}
 \label{pol_trans_coeff_def}
  D_{i' j} & = &
  \frac{ \frac{d \sigma}{ d \Omega_{1}^{'} 
  d E_{1}^{'}} ( \hat{s}_{i}, \hat{s}_{f} ) -
  \frac{ d \sigma}{ d \Omega_{1}^{'} d E_{1}^{'}}
  ( -\hat{s}_{i}, \hat{s}_{f} ) -
  \frac{ d \sigma}{ d \Omega_{1}^{'} d E_{1}^{'}}
  ( \hat{s}_{i}, -\hat{s}_{f} ) +
  \frac{ d \sigma}{ d \Omega_{1}^{'} d E_{1}^{'}}
  ( -\hat{s}_{i}, -\hat{s}_{f} ) }
  { \frac{d \sigma}{ d \Omega_{1}^{'} 
  d E_{1}^{'}} ( \hat{s}_{i}, \hat{s}_{f} ) + 
  \frac{ d \sigma}{ d \Omega_{1}^{'} d E_{1}^{'}}
  ( -\hat{s}_{i}, \hat{s}_{f} ) +
  \frac{ d \sigma}{ d \Omega_{1}^{'} d E_{1}^{'}}
  ( \hat{s}_{i}, -\hat{s}_{f} ) +
  \frac{ d \sigma}{ d \Omega_{1}^{'} d E_{1}^{'}}
  ( -\hat{s}_{i}, -\hat{s}_{f} ) }.
\end{eqnarray}
In Eq. (\ref{pol_trans_coeff_def}) a typical polarized differential cross
section is:
\begin{eqnarray}
 \label{double_dif_cross_section_dij_1}
  \frac{d \sigma }{d \Omega_{1}^{'} d E_{1}^{'} }
  ( \hat{s}_{i}, \hat{s}_{f} )
  & = &
  \frac{ | \vec{k}_{1} | E_{1}^{'} }{ | \vec{q} \, | E_{1}^{*^{'}}}
  \int_{p_{min}}^{k_{f}} \, d | \vec{p}_{2} \, \, | d \phi \, \, 
  | \vec{p}_{2} | f ( p_{1}^{*}, p_{2}^{*}, k_{f} )  
  \frac{1}{2} \tilde{\Gamma} ( \hat{s}_{i}, \hat{s}_{f} )
\end{eqnarray}
where
\begin{eqnarray*}
  \tilde{\Gamma} ( \vec{p}_{1}, ..., \vec{k}_{2},
  \hat{s}_{i}, \hat{s}_{f} )
  & = &
  \sum_{s_{2}, s_{2}^{'}} \, | {\cal M} |^{2}
  \\
  & = &
  g_{1}^{2} \sum_{m,n = 1}^{8} \, b_{m}^{*} b_{n} 
  [ {\rm Tr} ( \hat{P} ( \hat{s}_{i} ) \chi_{m}^{(1)} 
  \hat{P} ( \hat{s}_{f}) \chi_{n}^{(1)} ) ]
  [ {\rm Tr} ( \chi_{m}^{(2)} \chi_{n}^{(2)} ) ]
  \\
  \nonumber
  & = &
  f_{1} ( \vec{p}_{1}, ..., \vec{k}_{2} ) + 
  \vec{A}_{1} \cdot \hat{s}_{i} + 
  \vec{A}_{2} \cdot \hat{s}_{f} + 
  \\
  \nonumber
  & &
  ( \hat{s}_{i} \cdot \vec{A}_{3} ) 
  ( \hat{s}_{f} \cdot \vec{A}_{4} ) +
  ( \hat{s}_{i} \cdot \hat{s}_{f} ) ( \vec{A}_{6} \cdot \vec{A}_{7} ) +
  \hat{s}_{i} \cdot ( \hat{s}_{f} \times \vec{A}_{5} )
\end{eqnarray*}
with $ \vec{A}_{i} $ functions of only the three-momenta, $ \vec{p}_{1} $
to $ \vec{k}_{2} $ of which the explicit form can be inferred from 
Eq. (\ref{big_gamma_double_prime_xxx}).
Define again, now dictated by the form of 
Eq. (\ref{pol_trans_coeff_def}), a function:
\begin{eqnarray}
  \nonumber
  \Gamma ( \vec{p}_{1}, ..., \vec{k}_{2}, \hat{s}_{i},
  \hat{s}_{f} )
  & = &
  \tilde{\Gamma} ( \hat{s}_{i}, \hat{s}_{f} ) -
  \tilde{\Gamma} ( -\hat{s}_{i}, \hat{s}_{f} ) - 
  \tilde{\Gamma} ( \hat{s}_{i}, -\hat{s}_{f} ) +
  \\
 \nonumber
  & &
  \tilde{\Gamma} ( -\hat{s}_{i}, -\hat{s}_{f} ). 
  \\
 \nonumber
  & = &
  4 [ ( \hat{s}_{i} \cdot \vec{A}_{3} ) 
  ( \hat{s}_{f} \cdot \vec{A}_{4} ) + 
  \hat{s}_{i} \cdot ( \hat{s}_{f} \times \vec{A}_{5} ) +
  \\
 \label{big_gamma_1}
  & &
  ( \hat{s}_{i} \cdot \hat{s}_{f} ) 
  ( \vec{A}_{6} \cdot \vec{A}_{7} ) ].
\end{eqnarray}
The explicit expression for $ \Gamma $ contains 
various kinematical parameters
which are presented in the first column of  
Table~\ref{kin_for_spinobs_1}. The other columns contain 
the values of these quantities in the laboratory frame, 
for each polarization transfer observable:
$ \theta $ refers to the laboratory scattering angle,
$ p_{1} \, = \, | \vec{p}_{1} | $ and 
$ k_{1} \, = \, | \vec{k}_{1} | $.
Use of Eqs. (\ref{double_dif_cross_section_dij_1}),
(\ref{big_gamma_1}) and (\ref{unpol_cross_section}) lead to
\begin{eqnarray}
 \label{pol_trans_coeff_1}
  D_{i' j} & = &
  \frac{ \int_{p_{min}}^{k_{f}} d | \vec{p}_{2} | \, \, d \phi \, \, \, 
  | \vec{p}_{2} | f ( p_{1}^{*}, p_{2}^{*}, k_{f} ) \, \,
  4 \Gamma ( \hat{s}_{i}, \hat{s}_{f} ) }
  { \int_{p_{min}}^{k_{f}} d | \vec{p}_{2} | \, \, d \phi \, \, 
  | \vec{p}_{2} | f ( p_{1}^{*}, p_{2}^{*}, k_{f} ) \, \, 
  \Gamma^{''} ( \vec{p}_{1}, ..., \vec{k}_{2} ) }.
\end{eqnarray}
The polarization transfer observables for the 
$ ( \vec{p}, \vec{p}^{\, \, '} ) $ reaction are given by
{\small
\begin{eqnarray*}
  D_{i' j} [ ( \vec{p}, \vec{p}^{\, \, '} ) ]
  & = &
  \frac{ \int_{p_{min}}^{k_{f}} 
  d | \vec{p}_{2} | \, \, d \phi \, \, | \vec{p}_{2} |
  f ( k_{f} ) \,
  ( 4 Z_{eff} \Gamma ( \hat{s}_{i}, \hat{s}_{f}, 
  \{ b_{i} ( I = 1 ) \} ) \, + \, 
  4 N_{eff} \Gamma ( \hat{s}_{i}, \hat{s}_{f}, \{ b_{i}^{ave} \} ) )}
  { \int_{p_{min}}^{k_{f}} 
  d | \vec{p}_{2} | \, \, d \phi \, \, | \vec{p}_{2} |
  f ( k_{f} ) \,
  ( Z_{eff} \Gamma^{''} 
  ( \{ b_{i} ( I = 1 ) \} ) \, + \, 
  N_{eff} \Gamma^{''} 
  ( \{ b_{i}^{ave} \} ) )}
\end{eqnarray*}
}
and the corrsponding observables for the 
$ ( \vec{p}, \vec{n} ) $ reaction are
given by
\begin{eqnarray}
 \label{pol_trans_coeff_pn}
  D_{i' j} [ ( \vec{p}, \vec{n} ) ]
  & = &
  \frac{ \int_{p_{min}}^{k_{f}} 
  d | \vec{p}_{2} | \, \, d \phi \, \, | \vec{p}_{2} |
  f ( k_{f} ) \,
  4 \Gamma ( \hat{s}_{i}, \hat{s}_{f}, \{ b_{i}^{ch-ex} \} ) }
  { \int_{p_{min}}^{k_{f}} 
  d | \vec{p}_{2} | \, \, d \phi \, \, | \vec{p}_{2} |
  f ( k_{f} ) \,
  \Gamma^{''} 
  ( \{ b_{i}^{ch-ex} \} ) }.
\end{eqnarray}
Once again, as in Eq. (\ref{analyzing_power_pn}),
the effective number of neutrons does not appear in 
Eq. (\ref{pol_trans_coeff_pn}).
\end{enumerate}
Although the primary aim of this paper is to present the theoretical formalism
for calculating quasielastic proton-nucleus polarization transfer observables,
in the next section we give a brief glimpse of the predictive power of the
formalism by applying it to quasielastic
$^{40}$Ca$( \vec{p}, \vec{p}^{\; '} )$ scattering at 500 MeV 
A systematic study of the predictive power of the model, as well as a 
comparison to IA1-based predictions, will be presented in a future paper.

\section{Results}
\label{section_results}

Before presenting the results we mention the numerical 
checks that were performed to verify that the transformation 
from invariant amplitudes $ F_{n}^{ \{\rho\} }$,
to effective amplitudes $ b_{n} $ was carried out 
correctly and that the expressions for the spin observables 
in terms of the effective amplitudes are indeed correct. 
For $ M_{1} \; = \; M_{2} \; = \; M $ only subclass 
$ \hat{F}^{11} $ contributes to the invariant matrix element
and the IA2 representation 
is therefore equivalent to the SPVAT form of $ \hat{F} $. 
We therefore verified that our expressions for the spin observables 
in terms of the effective amplitudes give exactly 
the same numerical result as the corresponding expressions in 
Ref. \cite{HorowitzMurdock88} which contain only the 
five SPVAT amplitudes. This confirms that the transformation 
to effective amplitudes has been carried out correctly 
for only the SPVAT covariants. To verify the transformation 
for covariants $ K_{6} $ to $ K_{13} $, we derived 
expressions for the spin observables directly for each 
individual covariant $ K_{6} $ to $ K_{13} $. This 
involves traces over Dirac matrices (as opposed to 
the trace algebra involving Pauli matrices presented in this paper) 
and provides a non-trivial check for the transformation 
involving covariants $ K_{6} $ to $ K_{13} $. 
The fact that two independent ways give numerically 
the same result for all spin observables confirms the correctness of 
the transformation to effective amplitudes and the 
expressions for the spin observables derived in this paper.

The formalism in the previous sections is now applied to 
quasielastic $^{40}$Ca$(\vec{p}, \vec{p}^{\, \, '})$ scattering at an 
incident laboratory kinetic energy of 500 MeV
and a laboratory scattering
angle of $ 19^{\circ} $.
In the original calculation of 
Horowitz and Murdock in Ref. \cite{HorowitzMurdock88}, it was found that the 
use of an effective mass for both the projectile and target nucleons moved
the theoretical calculation closer to the data \cite{carey84}
and below the free mass 
calculation for $ A_{y} $. This was referred to as the quenching effect in 
the analyzing power and claimed to be a ''relativistic signature''.
In Ref. \cite{HorowitzMurdock88} the SPVAT 
parameterization of $ \hat{F} $ was used. Fig~\ref{fig_ca40_500pp19} shows the
results employing the IA2 representation of $ \hat{F} $. The solid line
represents the calculation using an effective mass for the projectile and
target nucleons with $ \frac{M_{1}}{M} \, = \, 0.892 $ and 
$ \frac{M_{2}}{M} \, = \, 0.817 $ taken from Table II in
Ref. \cite{hillhouse94} for $^{40}$Ca at $ T_{lab} \, = \, 500 \, MeV $. The
dashed line is the free mass calculation. 
The data are from Ref. \cite{carey84}.
We notice that the quenching effect 
in $ A_{y} $ is very small compared to Fig. 6 of Ref. \cite{HorowitzMurdock88}
over the entire energy range. The result is that the IA2 calculation does not
describe $ A_{y} $ as well as the IA1 calculation of
Ref. \cite{HorowitzMurdock88}. For the other observables, the effective mass
and the free mass calculation do equally well. This is in contrast to the
result in Ref. \cite{HorowitzMurdock88} where the $ D_{i'j} $'s 
only preferred a
free mass calculation.

\section{Summary}
\label{section_summary}

We have presented a theoretical formalism to calculate polarization transfer 
observables for quasielastic proton-nucleus scattering using a general
Lorentz invariant representation of the nucleon-nucleon scattering
matrix. In this way we avoid the ambiguities which are inherent in the
previously-used five-term representation (the $ SPVAT $ form) of
$ \hat{F} $. In the process we have derived an effective t-matrix, 
which is a $ 4 \times 4 $ matrix and therefore more familiar to
Nuclear Physics, but which still contains all the information coming
from the relativistic analysis. This necessitates the transformation
from the 44 invariant amplitudes to a set of eight effective 
amplitudes as well as the derivation of new expressions for the
spin observables in terms of the effective amplitudes. By staying 
within the framework of the Relativistic Plane Wave Impulse Approximation 
(with its many simplifying features)
\footnote{all of which are motivated by experimental data on the 
spin observables} and using a general Lorentz invariant representation
of $ \hat{F} $, allows us to do an unambiguous investigation of 
medium effects via quasielastic proton-nucleus scattering. The first 
application of the formalism to the reaction 
$^{40}$Ca$(\vec{p}, \vec{p}^{\, \, '})$ at $ T_{lab} \, = \, 500 \, MeV $ and
$ \theta_{lab} \, = \, 19^{\circ} $ shows that the IA2 representation of
$ \hat{F} $ does not lead to such strong medium effects in any of the 
spin observables, in contrast to the results in Ref. \cite{HorowitzMurdock88}
where the medium effect was most noticable in $ A_{y} $. There it was also 
found that the use of an effective mass for the projectile and target nucleons
led to the theoretical calculation being closer to the data than the free mass
calculation. The IA2 representation is consistent with data, however,
in that it 
predicts little medium effect in any of the spin observables, eventhough the
prediction of $ A_{y} $ is now a little poorer than before. In a subsequent
paper, a systematic study of spin observables, using the IA2 representation of
$ \hat{F} $, will be presented for both quasielastic 
$ (\vec{p}, \vec{p}^{\, \, '} ) $ and
$ ( \vec{p}, \vec{n} ) $ data.

\acknowledgements

One us (vdV) wishes to thank prof J.A. Tjon (University of Utrecht, the 
Netherlands) and prof S.J. Wallace (University of Maryland, USA) for very 
helpfull discussions. The financial assistance to vdV by the Harry Crossley 
Foundation, the South African FRD and the National Accelerator Centre 
is gratefully acknowledged.

\appendix
\section*{
Explicit expressions for spin observables in terms
of effective amplitudes $ \lowercase{a_{i} \, ( i \, = \, 1 - 8)} $.}
\label{spin_observables_xxx}

In this appendix we present explicit expressions for the quantities 
$ \Gamma^{''} $, $ \Gamma^{'} $ and $ \Gamma $ in terms of the effective 
amplitudes $ a_{i} $ which are related as follows to the effective amplitudes
$ b_{i} $:
\begin{eqnarray*}
 b_{1} & = & a_{i}
 \\
 b_{2} & = & \frac{a_{2}}{m^{4}}, 
 \\
 b_{i} & = & \frac{i}{m^{2}} a_{i} \qquad {\rm for} \, \, i = 3, 4
 \qquad {\rm and } 
 \\
 b_{i} & = & \frac{1}{m^{2}} a_{i} \qquad {\rm for} \, \, i = 5,6,7,8 
\end{eqnarray*}
where $ m $ denotes the free nucleon mass.

{\footnotesize
\begin{eqnarray}
\nonumber
\lefteqn{
\frac{1}{g_{1}^{2}}
\Gamma^{''} ( \vec{p}_{1}, \vec{p}_{2}, \vec{k}_{1},
\vec{k}_{2} ) = } 
\\ 
\nonumber
& &
4\,{{{ {\rm Im} ( a_{1} )}}^2} + 4\,{{{ {\rm Re} ( a_{1} )}}^2} + 
  {({{ \vec{N} \cdot \vec{N}}})^2}\,
\left( {{4\,{{{ {\rm Im} ( a_{2} )}}^2}}\over 
       {{m^8}}} + {{4\,{{{ {\rm Re} ( a_{2} )}}^2}}\over {{m^8}}}
      \right)  + 
\\
\nonumber
& &
{ \vec{N} \cdot \vec{N}}\,
   \left( {{4\,{{{ {\rm Im} ( a_{3} )}}^2}}\over {{m^4}}} + 
     {{4\,{{{ {\rm Re} ( a_{3} )}}^2}}\over {{m^4}}} + 
     {{4\,{{{ {\rm Im} ( a_{4} )}}^2}}\over {{m^4}}} + 
     {{4\,{{{ {\rm Re} ( a_{4} )}}^2}}\over {{m^4}}} \right)  + 
  \left( {{4\,{{{ {\rm Im} ( a_{6} )}}^2}}\over {{m^4}}} + 
     {{4\,{{{ {\rm Re} ( a_{6} )}}^2}}\over {{m^4}}} \right) \,
   {({{ \vec{p}_{a} \cdot \vec{p}_{a}})}^2} + 
\\
\nonumber
& &
\left( {{8\,{ {\rm Im} ( a_{6} )}\,{ {\rm Im} ( a_{7} )}}\over
         {{m^4}}} + {{8\,{ {\rm Re} ( a_{6} )}\,{ {\rm Re} ( a_{7} )}}\over 
       {{m^4}}} + {{8\,{ {\rm Im} ( a_{6} )}\,{ {\rm Im} ( a_{8} )}}\over 
       {{m^4}}} + {{8\,{ {\rm Re} ( a_{6} )}\,
{ {\rm Re} ( a_{8} )}}\over {{m^4}}}
      \right) \,{ \vec{p}_{a} \cdot 
\vec{p}_{a}}\,{ \vec{p}_{a} \cdot \vec{q}} + 
\\
\nonumber
& &
  \left( {{8\,{ {\rm Im} ( a_{5} )}\,{ {\rm Im} ( a_{6} )}}\over {{m^4}}} + 
     {{8\,{ {\rm Re} ( a_{5} )}\,{ {\rm Re} ( a_{6} )}}\over {{m^4}}} + 
     {{8\,{ {\rm Im} ( a_{7} )}\,{ {\rm Im} ( a_{8} )}}\over {{m^4}}} + 
     {{8\,{ {\rm Re} ( a_{7} )}\,{ {\rm Re} ( a_{8} )}}\over {{m^4}}} \right) \,
   {({{ \vec{p}_{a} \cdot \vec{q}} \; )}^2} + 
\\
\nonumber
& &
  \left( {{8\,{ {\rm Im} ( a_{5} )}\,{ {\rm Im} ( a_{7} )}}\over {{m^4}}} + 
     {{8\,{ {\rm Re} ( a_{5} )}\,{ {\rm Re} ( a_{7} )}}\over {{m^4}}} + 
     {{8\,{ {\rm Im} ( a_{5} )}\,{ {\rm Im} ( a_{8} )}}\over {{m^4}}} + 
     {{8\,{ {\rm Re} ( a_{5} )}\,{ {\rm Re} ( a_{8} )}}\over {{m^4}}} \right) \,
   { \vec{p}_{a}\cdot \vec{q}}\,{ \vec{q}\cdot \vec{q}} + 
\\
\label{big_gamma_double_prime_xxx}
& &
  \left( {{4\,{{{ {\rm Im} ( a_{5} )}}^2}}\over {{m^4}}} + 
     {{4\,{{{ {\rm Re} ( a_{5} )}}^2}}\over {{m^4}}} \right) \,
   {({{ \vec{q}\cdot \vec{q}})}^2} +
\left( {{4\,{{{ {\rm Im} ( a_{7} )}}^2}}\over 
       {{m^4}}} + {{4\,{{{ {\rm Re} ( a_{7} )}}^2}}\over {{m^4}}} + 
     {{4\,{{{ {\rm Im} ( a_{8} )}}^2}}\over {{m^4}}} + 
     {{4\,{{{ {\rm Re} ( a_{8} )}}^2}}\over {{m^4}}} \right) \,
   { \vec{p}_{a} \cdot \vec{p}_{a}}\,{ \vec{q}\cdot \vec{q}} 
\end{eqnarray}

\begin{eqnarray}
\nonumber
\lefteqn{
\frac{1}{g_{1}^{2}} \Gamma^{'} ( \vec{p}_{1}, ..., \vec{k}_{2},
\hat{s}_{f} ) =}
\\
\nonumber
& &
\left( {{-4\,{ {\rm Re} ( a_{2} )}\,{ {\rm Im} ( a_{4} )}\,
{ \vec{N} \cdot \vec{N}}}\over 
       {{m^6}}} + {{4\,{ {\rm Im} ( a_{2} )}\,{ {\rm Re} ( a_{4} )}\,
         { \vec{N} \cdot \vec{N}}}\over {{m^6}}} - 
     {{4\,{ {\rm Re} ( a_{1} )}\,{ {\rm Im} ( a_{3} )}}\over {{m^2}}} + 
     {{4\,{ {\rm Im} ( a_{1} )}\,{ {\rm Re} ( a_{3} )}}\over {{m^2}}} \right) \,
   { \vec{N} \cdot \hat{s}_{f}} + 
\\
\nonumber
& &
{ \vec{p}_{a} \cdot ( \vec{q} \times \hat{s}_{f} )}\,
   \left( {{4\,{ {\rm Re} ( a_{6} )}\,{ {\rm Im} ( a_{7} )}\,
   { \vec{p}_{a} \cdot 
   \vec{p}_{a}}}\over 
       {{m^4}}} - {{4\,{ {\rm Im} ( a_{6} )}\,{ {\rm Re} ( a_{7} )}\,
         { \vec{p}_{a} \cdot \vec{p}_{a}}}\over {{m^4}}} - 
     {{4\,{ {\rm Re} ( a_{5} )}\,{ {\rm Im} ( a_{6} )}\,{ \vec{p}_{a} \cdot 
     \vec{q}}}\over {{m^4}}} + 
     {{4\,{ {\rm Im} ( a_{5} )}\,{ {\rm Re} ( a_{6} )}\,{ \vec{p}_{a} \cdot 
     \vec{q}}}\over {{m^4}}} -
\right.
\\
\label{big_gamma_prime_xxx}
& &
\left.
     {{4\,{ {\rm Re} ( a_{7} )}\,{ {\rm Im} ( a_{8} )}\,{ \vec{p}_{a} \cdot 
     \vec{q}}}\over {{m^4}}} + 
     {{4\,{ {\rm Im} ( a_{7} )}\,{ {\rm Re} ( a_{8} )}\,{ \vec{p}_{a} \cdot 
     \vec{q}}}\over {{m^4}}} - 
     {{4\,{ {\rm Re} ( a_{5} )}\,{ {\rm Im} ( a_{8} )}\,{ \vec{q} \cdot 
     \vec{q}}}\over {{m^4}}} + 
     {{4\,{ {\rm Im} ( a_{5} )}\,{ {\rm Re} ( a_{8} )}\,{ \vec{q} \cdot 
     \vec{q}}}\over {{m^4}}} \, \, 
      \right) 
\end{eqnarray}

\begin{eqnarray}
\nonumber
\lefteqn{
\frac{1}{4 g_{1}^{2}} \,
\Gamma ( \vec{p}_{1}, ..., \vec{k}_{2},
\hat{s}_{i}, \hat{s}_{f} ) =}
\\
\nonumber
& &
\left( {{2\,{{{ {\rm Im} ( a_{2} )}}^2}\,{ \vec{N} \cdot \vec{N}}}\over 
 {{m^8}}} + {{2\,{{{ {\rm Re} ( a_{2} )}}^2}\,{ \vec{N} \cdot \vec{N}}}\over
         {{m^8}}} + 
{{2\,{{{ {\rm Im} ( a_{3} )}}^2}}\over {{m^4}}} + 
     {{2\,{{{ {\rm Re} ( a_{3} )}}^2}}\over {{m^4}}} \right) \,
   { \vec{N} \cdot \hat{s}_{f}}\,{ \vec{N} \cdot \hat{s}_{i}} + 
  \left( {{2\,{ {\rm Im} ( a_{2} )}\,
{ {\rm Im} ( a_{4} )}\,{ \vec{N} \cdot \vec{N}}}\over 
       {{m^6}}} + 
\right.
\\
\nonumber
& &
\left.
{{2\,{ {\rm Re} ( a_{2} )}\,{ {\rm Re} ( a_{4} )}\,
         { \vec{N} \cdot \vec{N}}}\over {{m^6}}} - 
     {{2\,{ {\rm Im} ( a_{1} )}\,{ {\rm Im} ( a_{3} )}}\over {{m^2}}} - 
     {{2\,{ {\rm Re} ( a_{1} )}\,{ {\rm Re} ( a_{3} )}}\over {{m^2}}} \right) \,
   { \vec{N} \cdot ( \hat{s}_{i} \times \hat{s}_{f} )} + 
{ \vec{p}_{a} \cdot \hat{s}_{f}}\,{ \vec{p}_{a} \cdot \hat{s}_{i}}\,
   \left( {{2\,{{{ {\rm Im} ( a_{6} )}}^2}\,{ \vec{p}_{a} 
         \cdot \vec{p}_{a}}}\over {{m^4}}} + 
\right.
\\
\nonumber
& &
\left.
{{2\,{{{ {\rm Re} ( a_{6} )}}^2}\,{ \vec{p}_{a} \cdot \vec{p}_{a}}}\over {{m^4}}} +
{{4\,{ {\rm Im} ( a_{6} )}\,{ {\rm Im} ( a_{8} )}\,{ \vec{p}_{a} \cdot \vec{q}}}\over {{m^4}}} + 
{{4\,{ {\rm Re} ( a_{6} )}\,{ {\rm Re} ( a_{8} )}\,{ \vec{p}_{a} \cdot \vec{q}}}\over {{m^4}}} + 
{{2\,{{{ {\rm Im} ( a_{8} )}}^2}\,{ \vec{q} \cdot \vec{q}}}\over {{m^4}}} +
{{2\,{{{ {\rm Re} ( a_{8} )}}^2}\,
{ \vec{q} \cdot \vec{q}}}\over {{m^4}}} \right) +
\\
\nonumber
& &
{ \vec{p}_{a} \cdot \hat{s}_{i}}\,\left( {{2\,{ {\rm Im} ( a_{6} )}\,{ {\rm Im} ( a_{7} )}\,
{ \vec{p}_{a} \cdot \vec{p}_{a}}}\over {{m^4}}} + 
{{2\,{ {\rm Re} ( a_{6} )}\,{ {\rm Re} ( a_{7} )}\,{ \vec{p}_{a} \cdot \vec{p}_{a}}}\over {{m^4}}} + 
{{2\,{ {\rm Im} ( a_{5} )}\,{ {\rm Im} ( a_{6} )}\,{ \vec{p}_{a} \cdot \vec{q}}}\over {{m^4}}} + 
{{2\,{ {\rm Re} ( a_{5} )}\,{ {\rm Re} ( a_{6} )}\,
{ \vec{p}_{a} \cdot \vec{q}}}\over {{m^4}}} +
\right.
\\
\nonumber
& &
\left.
{{2\,{ {\rm Im} ( a_{7} )}\,{ {\rm Im} ( a_{8} )}\,{ \vec{p}_{a} \cdot \vec{q}}}\over {{m^4}}} + 
{{2\,{ {\rm Re} ( a_{7} )}\,{ {\rm Re} ( a_{8} )}\,{ \vec{p}_{a} \cdot \vec{q}}}\over {{m^4}}} + 
{{2\,{ {\rm Im} ( a_{5} )}\,{ {\rm Im} ( a_{8} )}\,
{ \vec{q} \cdot \vec{q}}}\over {{m^4}}} + 
     {{2\,{ {\rm Re} ( a_{5} )}\,{ {\rm Re} ( a_{8} )}\,{ \vec{q} \cdot \vec{q}}}\over {{m^4}}}
      \right) \,{ \vec{q} \cdot \hat{s}_{f}} + 
\\
\nonumber
& &
{ \vec{p}_{a} \cdot \hat{s}_{f}}\,\left( {{2\,{ {\rm Im} ( a_{6} )}\,{ {\rm Im} ( a_{7} )}\,
{ \vec{p}_{a} \cdot \vec{p}_{a}}}\over {{m^4}}} + 
{{2\,{ {\rm Re} ( a_{6} )}\,{ {\rm Re} ( a_{7} )}\,{ \vec{p}_{a} \cdot \vec{p}_{a}}}\over {{m^4}}} + 
{{2\,{ {\rm Im} ( a_{5} )}\,{ {\rm Im} ( a_{6} )}\,{ \vec{p}_{a} \cdot \vec{q}}}\over {{m^4}}} + 
{{2\,{ {\rm Re} ( a_{5} )}\,{ {\rm Re} ( a_{6} )}\,
{ \vec{p}_{a} \cdot \vec{q}}}\over {{m^4}}} +
\right.
\\
\nonumber
& &
\left.
{{2\,{ {\rm Im} ( a_{7} )}\,{ {\rm Im} ( a_{8} )}\,{ \vec{p}_{a} \cdot \vec{q}}}\over {{m^4}}} + 
{{2\,{ {\rm Re} ( a_{7} )}\,{ {\rm Re} ( a_{8} )}\,{ \vec{p}_{a} \cdot \vec{q}}}\over {{m^4}}} + 
{{2\,{ {\rm Im} ( a_{5} )}\,{ {\rm Im} ( a_{8} )}\,{ \vec{q} \cdot \vec{q}}}\over {{m^4}}} + 
{{2\,{ {\rm Re} ( a_{5} )}\,{ {\rm Re} ( a_{8} )}\,{ \vec{q} \cdot \vec{q}}}\over {{m^4}}}
      \right) \,{ \vec{q} \cdot \hat{s}_{i}} + 
\\
\nonumber
& &
\left( {{2\,{{{ {\rm Im} ( a_{7} )}}^2}\,{ \vec{p}_{a} \cdot \vec{p}_{a}}}\over {{m^4}}} + 
     {{2\,{{{ {\rm Re} ( a_{7} )}}^2}\,{ \vec{p}_{a} \cdot \vec{p}_{a}}}\over {{m^4}}} + 
     {{4\,{ {\rm Im} ( a_{5} )}\,{ {\rm Im} ( a_{7} )}\,{ \vec{p}_{a} \cdot \vec{q}}}\over {{m^4}}} + 
     {{4\,{ {\rm Re} ( a_{5} )}\,{ {\rm Re} ( a_{7} )}\,{ \vec{p}_{a} \cdot \vec{q}}}\over {{m^4}}} +
{{2\,{{{ {\rm Im} ( a_{5} )}}^2}\,{ \vec{q} \cdot \vec{q}}}\over {{m^4}}} +
\right.
\\
\nonumber
& &
\left.
{{2\,{{{ {\rm Re} ( a_{5} )}}^2}\,{ \vec{q} \cdot \vec{q}}}\over {{m^4}}} \right) 
\,{ \vec{q} \cdot \hat{s}_{f}}\,{ \vec{q} \cdot \hat{s}_{i}} + 
\left( {{{ {\rm Im} ( a_{1} )}}^2} + {{{ {\rm Re} ( a_{1} )}}^2} - 
{{{{{ {\rm Im} ( a_{2} )}}^2}\,{({{ \vec{N} \cdot \vec{N}})}^2}}\over 
{{m^8}}} - {{{{{ {\rm Re} ( a_{2} )}}^2}\,
{({{ \vec{N} \cdot \vec{N}})}^2}}\over {{m^8}}} - 
{{{{{ {\rm Im} ( a_{3} )}}^2}\,{ \vec{N} \cdot \vec{N}}}\over {{m^4}}} - 
\right.
\\
\nonumber
& &
\left.
{{{{{ {\rm Re} ( a_{3} )}}^2}\,{ \vec{N} \cdot \vec{N}}}\over {{m^4}}} + 
{{{{{ {\rm Im} ( a_{4} )}}^2}\,{ \vec{N} \cdot \vec{N}}}\over {{m^4}}} + 
{{{{{ {\rm Re} ( a_{4} )}}^2}\,{ \vec{N} \cdot \vec{N}}}\over {{m^4}}} - 
{{{{{ {\rm Im} ( a_{6} )}}^2}\,
{({{ \vec{p}_{a} \cdot \vec{p}_{a}})}^2}}\over {{m^4}}} - 
{{{{{ {\rm Re} ( a_{6} )}}^2}\,{({{ \vec{p}_{a} 
\cdot \vec{p}_{a}})}^2}}\over {{m^4}}} - 
\right.
\\
\nonumber
& &
\left.
{{2\,{ {\rm Im} ( a_{6} )}\,{ {\rm Im} ( a_{7} )}\,{ \vec{p}_{a} 
\cdot \vec{p}_{a}}\,{ \vec{p}_{a} \cdot \vec{q}}}\over 
{{m^4}}} - {{2\,{ {\rm Re} ( a_{6} )}\,{ {\rm Re} ( a_{7} )}\,{ \vec{p}_{a} \cdot \vec{p}_{a}}\,
{ \vec{p}_{a} \cdot \vec{q}}}\over {{m^4}}} - 
     {{2\,{ {\rm Im} ( a_{6} )}\,{ {\rm Im} ( a_{8} )}\,{ \vec{p}_{a} \cdot 
  \vec{p}_{a}}\,{ \vec{p}_{a} \cdot \vec{q}}}\over 
       {{m^4}}} - 
\right.
\\
\nonumber
& &
\left.
{{2\,{ {\rm Re} ( a_{6} )}\,{ {\rm Re} ( a_{8} )}\,{ \vec{p}_{a} \cdot \vec{p}_{a}}\,
         { \vec{p}_{a} \cdot \vec{q}}}\over {{m^4}}} - 
     {{2\,{ {\rm Im} ( a_{5} )}\,{ {\rm Im} ( a_{6} )}\,{({{ \vec{p}_{a} \cdot \vec{q}})}^2}}\over 
       {{m^4}}} - {{2\,{ {\rm Re} ( a_{5} )}\,{ {\rm Re} ( a_{6} )}\,
         {({{ \vec{p}_{a} \cdot \vec{q}})}^2}}\over {{m^4}}} - 
     {{2\,{ {\rm Im} ( a_{7} )}\,{ {\rm Im} ( a_{8} )}\,{({{ \vec{p}_{a} \cdot \vec{q}})}^2}}\over 
       {{m^4}}} - 
\right.
\\
\nonumber
& &
\left.
{{2\,{ {\rm Re} ( a_{7} )}\,{ {\rm Re} ( a_{8} )}\,
         {({{ \vec{p}_{a} \cdot \vec{q}})}^2}}\over {{m^4}}} - 
     {{{{{ {\rm Im} ( a_{7} )}}^2}\,{ \vec{p}_{a} \cdot 
    \vec{p}_{a}}\,{ \vec{q} \cdot \vec{q}}}\over 
       {{m^4}}} - {{{{{ {\rm Re} ( a_{7} )}}^2}\,{ \vec{p}_{a} \cdot \vec{p}_{a}}\,
         { \vec{q} \cdot \vec{q}}}\over {{m^4}}} - 
     {{{{{ {\rm Im} ( a_{8} )}}^2}\,{ \vec{p}_{a} \cdot 
     \vec{p}_{a}}\,{ \vec{q} \cdot \vec{q}}}\over 
{{m^4}}} - {{{{{ {\rm Re} ( a_{8} )}}^2}\,{ \vec{p}_{a} \cdot \vec{p}_{a}}\,
{ \vec{q} \cdot \vec{q}}}\over {{m^4}}} - 
\right.
\\
\nonumber
& &
\left.
     {{2\,{ {\rm Im} ( a_{5} )}\,{ {\rm Im} ( a_{7} )}\,{ \vec{p}_{a} \cdot 
     \vec{q}}\,{ \vec{q} \cdot \vec{q}}}\over 
       {{m^4}}} - {{2\,{ {\rm Re} ( a_{5} )}\,{ {\rm Re} ( a_{7} )}\,{ \vec{p}_{a} \cdot \vec{q}}\,
         { \vec{q} \cdot \vec{q}}}\over {{m^4}}} - 
     {{2\,{ {\rm Im} ( a_{5} )}\,{ {\rm Im} ( a_{8} )}\,{ \vec{p}_{a} 
     \cdot \vec{q}}\,{ \vec{q} \cdot \vec{q}}}\over 
       {{m^4}}} - {{2\,{ {\rm Re} ( a_{5} )}\,{ {\rm Re} ( a_{8} )}\,{ \vec{p}_{a} \cdot \vec{q}}\,
         { \vec{q} \cdot \vec{q}}}\over {{m^4}}} - 
\right.
\\
\label{big_gamma_xxx}
& &
\left.
{{{{{ {\rm Im} ( a_{5} )}}^2}\,{({{ \vec{q} \cdot \vec{q}})}^2}}
\over {{m^4}}} - 
{{{{{ {\rm Re} ( a_{5} )}}^2}\,{({{ \vec{q} \cdot \vec{q}})}^2}}\over {{m^4}}}
\right) \,{ \hat{s}_{i} \cdot \hat{s}_{f}}
\end{eqnarray}
}

\begin{figure}
\caption{
Two-body scattering process with momentum, mass and spin labels for
the external nucleons. $ \hat{F} $ is the $ 16 \times 16 $ 
nucleon-nucleon scattering matrix in the two-nucleon spin space. 
$ p_{i} $ and $ k_{i} $ $ ( i \, = \, 1 -- 2 ) $ represent the four-momenta
of the particles respectively. $ M_{1} $ and $ M_{2} $ denote the effective 
masses of the projectile and target nucleons respectively. 
$ s_{i} $, $ s_{2} $, $ s_{f} $ and $ s_{2}^{ \; '} $ are the 
spin four-vectors for each particle.
\label{two-body_scattering_process}}
\end{figure}   

\begin{figure}
\caption{Spin observables for a range of transferred energy $ \omega $ over 
the quasielastic peak for inclusive proton scattering from
$^{40}$Ca at 500 MeV and $ \theta_{lab} \, = \, 19^{\circ} $. The centroid
of the quasielastic peak is at $ \omega \, \approx \, 63 $ MeV.
Data are from Ref. [23]. The solid represents the IA2 calculation 
and the dashed line represents the free mass  calculation.
\label{fig_ca40_500pp19}}
\end{figure}   

\begin{table}
\begin{center}
\caption{Expressions for kinematical quantities containing 
$ \hat{s}_{i} $ and/or $ \hat{s}_{f} $ for each non-zero polarization
transfer observable.}
\label{kin_for_spinobs_1}
\begin{tabular}{|l|c|c|c|c|c|} \hline
 kinematical & & & & &
 \\
 quantity & $ D_{l' l} $ & $ D_{s' s} $ & $ D_{n n} $ &
 $ D_{s' l} $ & $ D_{l' s} $
 \\[3mm]
 \hline
 $ \vec{q} \cdot \hat{s}_{i} $ & $ p_{1} - k_{1} \cos \theta $ &
 $ -k_{1} \sin \theta $ & 0 & $ p_{1} - k_{1} \cos \theta $ &
 $ -k_{1} \sin \theta $
 \\[3mm]
 \hline
 $ \vec{q} \cdot \hat{s}_{f} $ & $ p_{1} \cos \theta - k_{1} $ &
 $ -p_{1} \sin \theta $ & 0 & $ -p_{1} \sin \theta $ & 
 $ p_{1} \cos \theta - k_{1} $
 \\[3mm]
 \hline
 $ \vec{p}_{a} \cdot \hat{s}_{i} $ & 
 $ \frac{1}{2} ( p_{1} + k_{1} \cos \theta ) $ &
 $ \frac{1}{2} k_{1} \sin \theta $ & 0 &
 $ \frac{1}{2} ( p_{1} + k_{1} \cos \theta ) $ &
 $ \frac{1}{2} k_{1} \sin \theta $
 \\[3mm]
 \hline
 $ \vec{p}_{a} \cdot \hat{s}_{f} $ & 
 $ \frac{1}{2} ( p_{1} \cos \theta + k_{1} ) $ &
 $ -\frac{1}{2} p_{1} \sin \theta $ & 0 &
 $ -\frac{1}{2} p_{1} \sin \theta $ &
 $ \frac{1}{2} ( p_{1} \cos \theta + k_{1} ) $ 
 \\[3mm]
 \hline
 $ \vec{N} \cdot \hat{s}_{i} $ & 0 & 0 & 
 $ p_{1} k_{1} \sin \theta $ & 0 & 0
 \\[3mm]
 \hline
 $ \vec{N} \cdot \hat{s}_{f} $ & 0 & 0 & 
 $ p_{1} k_{1} \sin \theta $ & 0 & 0
 \\[3mm]
 \hline
 $ \hat{s}_{i} \cdot \hat{s}_{f} $ & $ \cos \theta $ & 
 $ \cos \theta $ & 1 & $ -\sin \theta $ & $ \sin \theta $ 
 \\[3mm]
 \hline
 $ \vec{N} \cdot ( \hat{s}_{i} \times \hat{s}_{f} ) $ &
 $ p_{1} k_{1} \sin^{2} \theta $ & 
 $ p_{1} k_{1} \sin^{2} \theta $ & 0 & 
 $ p_{1} k_{1} \cos \theta \sin \theta $ &
 $ -p_{1} k_{1} \cos \theta \sin \theta $
 \\
 \hline
\end{tabular}
\end{center}
\end{table}

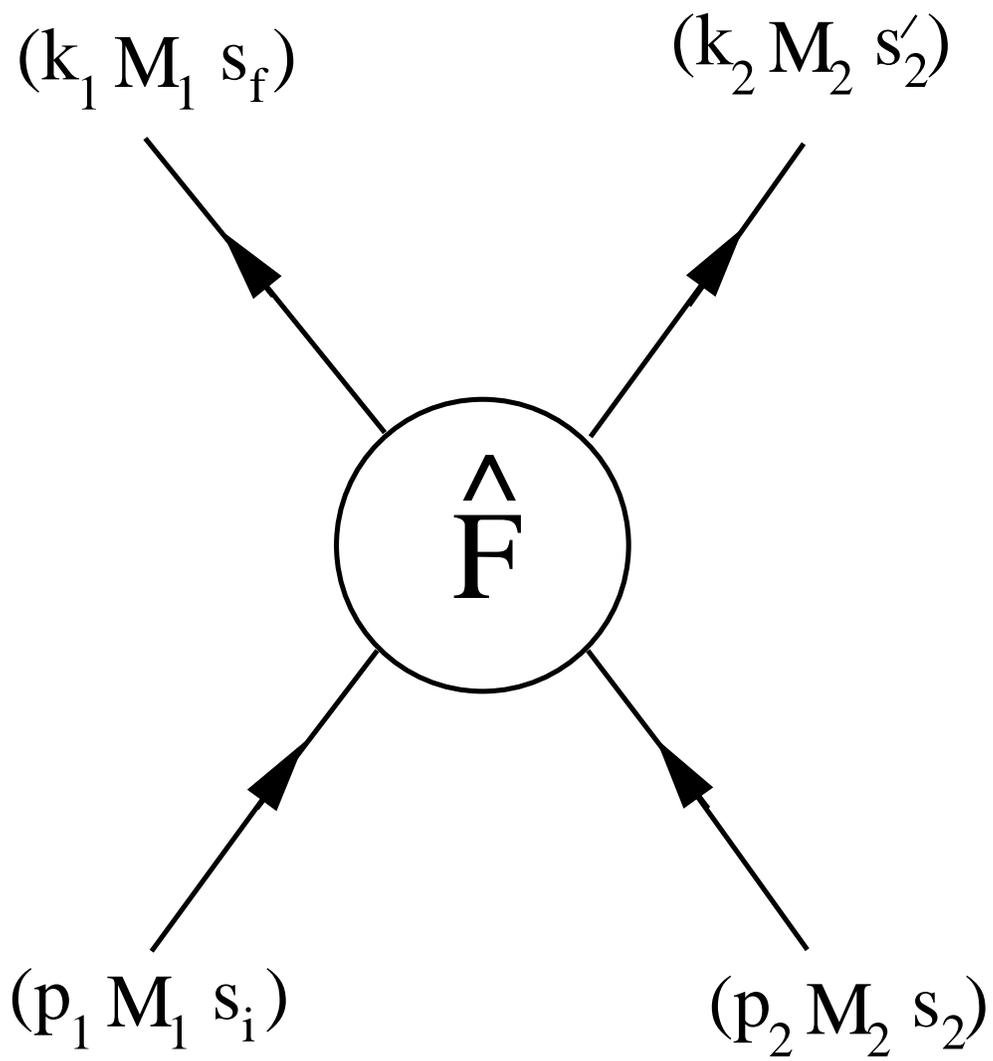
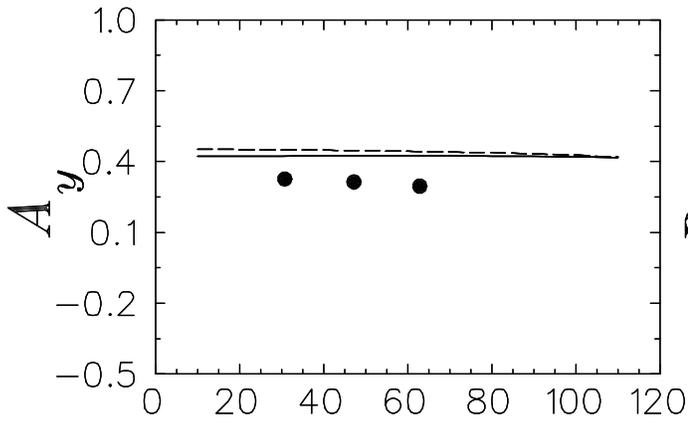
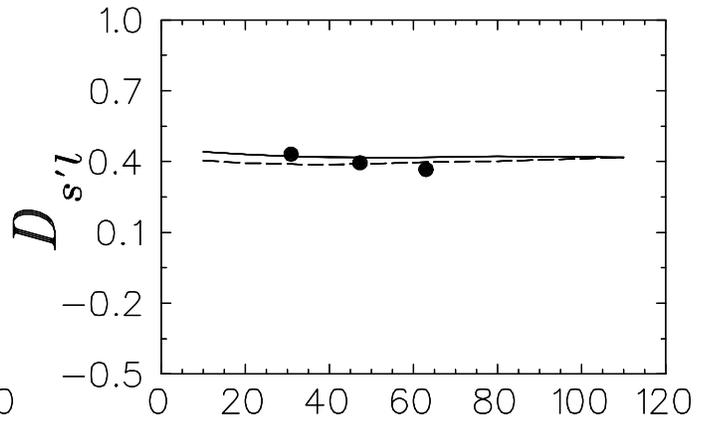
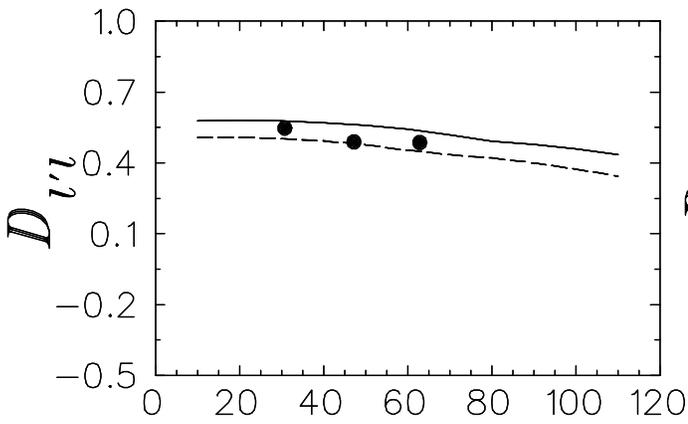
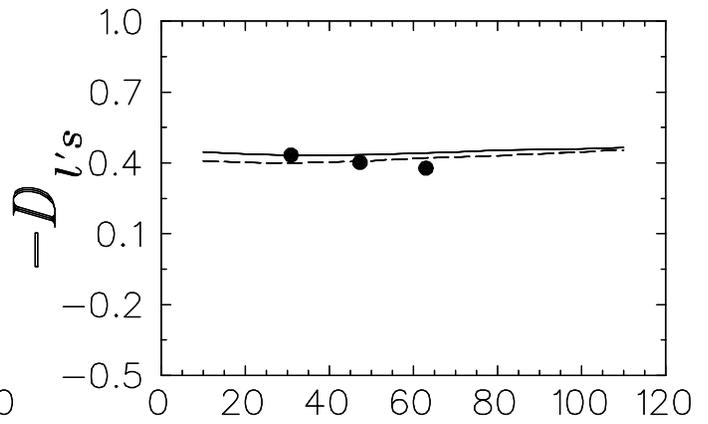
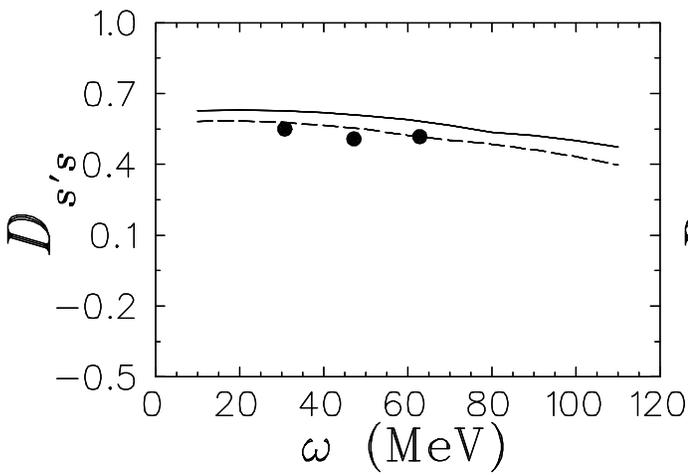
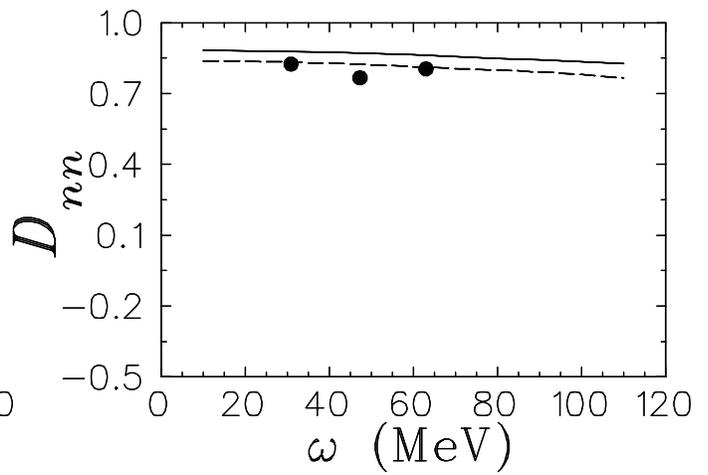
\end{document}